\documentclass[twocolumn,aps,preprintnumbers,amsmath,amssymb]{revtex4-2} 
\usepackage{graphicx} 
\usepackage{hyperref} 
\usepackage{amsmath} 
\usepackage{amssymb} 
\usepackage{float} 
\usepackage{caption}
\usepackage{subcaption} 
\usepackage{placeins} 
\usepackage{nameref}

\begin{document}

\newcommand{\inlineheading}[1]{\vspace{0.5em}\noindent\textbf{#1:}\quad\vspace{0.5em}}

\title{IEC-Independent Coupling Between Water Uptake and Ionic Conductivity in Anion-Conducting Polymer Films}

\author{Joan M. Montes de Oca}
\thanks{These authors contributed equally to this work}
\affiliation{Pritzker School of Molecular Engineering, University of Chicago, Chicago, Illinois 60637, United States}

\author{Ruilin Dong}
\thanks{These authors contributed equally to this work}
\affiliation{Pritzker School of Molecular Engineering, University of Chicago, Chicago, Illinois 60637, United States}

\author{Zhongyang Wang}
\affiliation{Department of Chemical \& Biological Engineering, The University of Alabama, Tuscaloosa, Alabama 35487-0203, United States}

\author{Gervasio Zaldívar}
\affiliation{Department of Chemical and Biomolecular Engineering, Tandon School of Engineering, New York University, Brooklyn, NY, USA}

\author{Ge Sun}
\affiliation{Department of Chemical and Biomolecular Engineering, Tandon School of Engineering, New York University, Brooklyn, NY, USA}

\author{Shrayesh N. Patel}
\affiliation{Pritzker School of Molecular Engineering, University of Chicago, Chicago, Illinois 60637, United States}

\author{Paul F. Nealey}
\affiliation{Pritzker School of Molecular Engineering, University of Chicago, Chicago, Illinois 60637, United States}
\email{nealey@uchicago.edu}

\author{Juan J. de Pablo}
\affiliation{Department of Chemical and Biomolecular Engineering, Tandon School of Engineering, New York University, Brooklyn, NY, USA}

\email{depablo@uchicago.edu}

\date{\today}

\begin{abstract}
Anion exchange membranes (AEMs) are promising candidates for replacing proton exchange membranes (PEMs) in electrochemical devices such as fuel cells, electrolyzers, batteries, and osmotic energy extraction systems. However, optimizing the AEM design requires a deeper understanding of the ionic conduction mechanism in the hydrated polymer matrix. This study investigates this mechanism by seeking to understand the relationship between ion exchange capacity (IEC), water absorption, and ionic conductivity in polynorbornene-based thin films. We combine experimental measurements with computational simulations using a newly developed minimal model of the polymer film. Our model is able to reproduce key experimental observations, including water sorption isotherms and ion conduction behavior as a function of relative humidity, and successfully captures the relationship between them.
By comparing experimental data with computational results, we explain the commonly reported correlation between conductivity and hydration level and show how the correlation between these variables is affected by the charge density and temperature of the material. Our research advances our understanding of the physical mechanisms that govern the performance of the polyelectrolyte membrane, which is essential for the development of more efficient, stable, and environmentally friendly electrochemical systems.

\end{abstract}
\maketitle

\section{Introduction}

Ionic Exchange Membranes (IEMs) are fundamental components in various electrochemical devices like fuel cells, electrolyzers, and batteries~\cite{AEM_electro_review,intro_review_AEM,review_AEM_chem,intro_electro_CO2}. Their primary function is to selectively and efficiently transport specific ions while restricting the crossover of unwanted species, such as fuel and oxidant, which is critical for maintaining efficiency and performance. The performance of IEMs can be characterized by their ionic conductivity and water uptake properties, which are influenced by the structure and chemistry of the membrane. Understanding these properties is important for the development of more efficient and durable electrochemical systems.

Structurally, IEMs consist of a hydrophobic polymer backbone functionalized with charged side chains. In anion exchange membranes (AEMs), the polymer backbone bears positively charged groups, which are compensated by mobile anions. 

Proton exchange membranes (PEMs) have historically monopolized scientific, technological, and commercial efforts in the advancement of IEMs. However, despite the high performance of PEMs, the industry is now shifting its focus towards developing IEMs that can conduct ions other than protons.  This transition aims to replace costly platinum electrodes in fuel cells, achieve greater stability in alkaline environments, and develop chemistries that are more environmentally friendly than the fluorinated compounds typically found in commercial PEMs. One promising alternative to PEMs is anion exchange membranes (AEMs).

However, to develop AEMs with the electrical performance and chemical and mechanical stability that modern applications require, it is necessary to have a detailed understanding of how typical design variables (e.g. ion exchange capacity and polymer architecture) impact these properties.~\cite{chris_arges_AEM,AEM_electro_review,AMEWS_review} In particular, the relationship between the hydration level of the AEM and its ionic conductivity is not fully understood despite playing a central role in optimizing the design of IEM. However, it is known that a higher degree of hydration is correlated with higher ionic conductivity, often at the expense of the mechanical properties of the material~\cite{water_degrades_mec}.

This work investigates the relationship between conductivity and water content in anion-conducting polynorbornene-based polyelectrolytes. Unlike conventional studies that focus on micrometer-scale membranes, we examine nanometer-scale films ($\sim 100\,\text{nm}$), allowing us to probe ion transport in relatively confined geometries. Polynorbornenes serve as an ideal polymer matrix due to their electrochemically stable hydrocarbon backbone and tunable functionalization~\cite{review_AEM_chem}. A key variable in our study is the ion exchange capacity (IEC), which defines the fixed charge concentration in the dry polymer and strongly influences water uptake and ion transport.

We analyze conductivity, water content, and ion transport from a bottom-up perspective by combining experiments with a newly developed computational model. Previously, molecular dynamics (MD) simulations provided insight into how conduction mechanisms evolve with hydration in AEMs~\cite{Zhongyang_Ge_AEM}. However, the high computational cost of MD limits the systematic exploration of the entire design space for AEMs. To overcome this, we introduce the Coupled Layers Model (CLM), a coarse-grained simulation framework that integrates two established approaches: (i) treating conduction as a random resistor network~\cite{kirkpatrick1973percolation,weber_kusoglu_3D_network,book_introduction_percolation}, and (ii) modeling water sorption using lattice-gas Monte Carlo simulations~\cite{monson2001Lattice_condensation,Lattice_adsorption_aranovich_1,Lattice_adsorption_aranovich_2}. By bridging these two physical processes, the CLM captures the interplay between hydration and ionic transport with significantly reduced computational cost.

Through experimental validation of the CLM, we aim to establish a robust framework for understanding the coupled effects of charge density, water uptake, and percolation on ionic conductivity. Our findings not only clarify the role of polymer chemistry in ion transport but also provide a scalable modeling strategy for systematic AEM design. This combined experimental and theoretical approach lays the groundwork for future studies exploring additional design parameters that govern membrane performance.

\subsection{Experimental Overview}

\begin{figure}[htbp]
    \centering
    \includegraphics[width=0.45\textwidth]{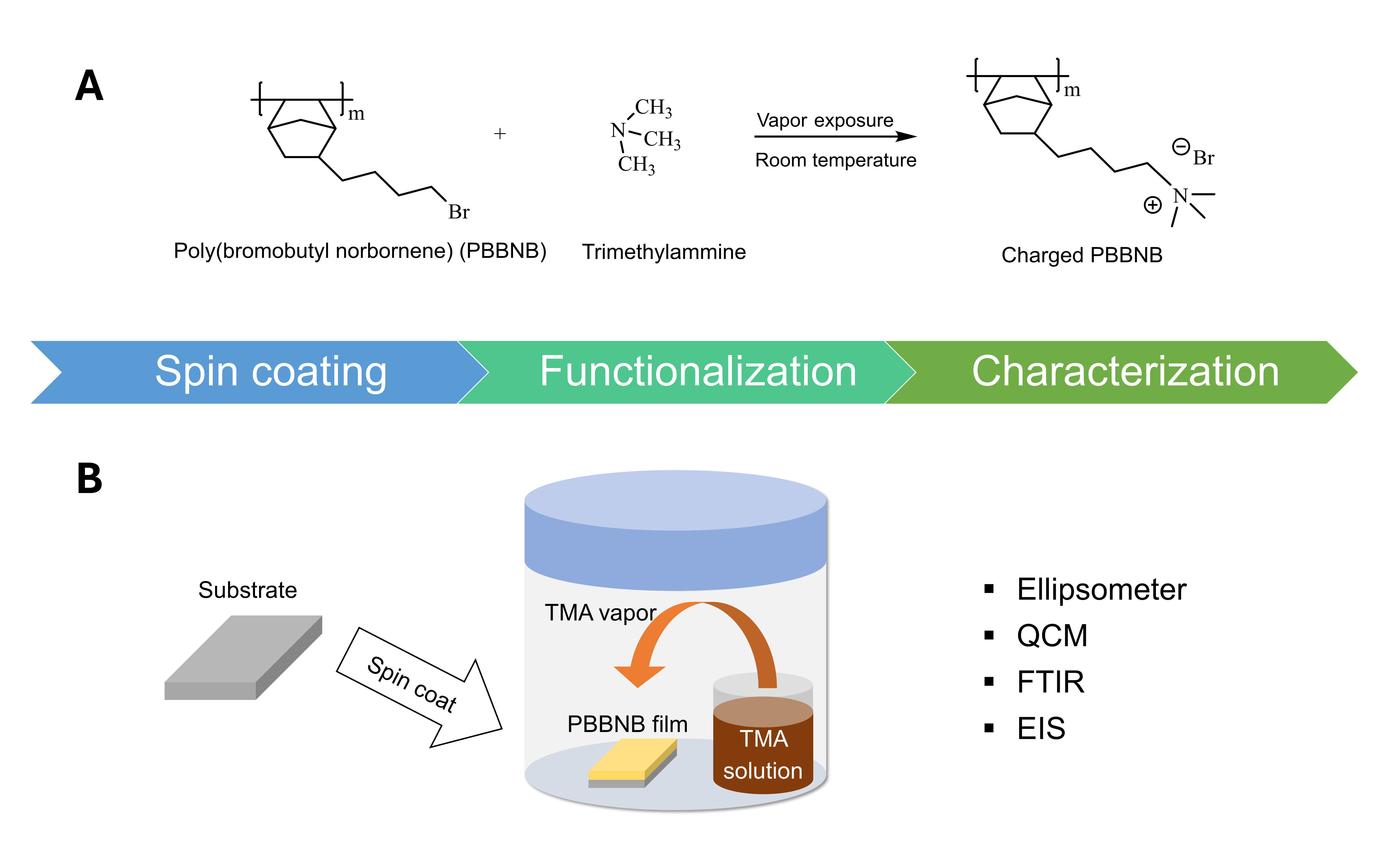}
    \caption{(A) Schematic of the TMA vapor infiltration reaction used to functionalize PBBNB thin films. (B) Overview of the experimental process: PBBNB thin films were fabricated via spin-coating onto different substrates, selected based on the required characterization techniques. The films were then functionalized in a sealed jar with saturated TMA vapor, converting bromide groups into quaternary ammonium salts. The ion exchange capacity (IEC) and degree of functionalization (DoF) were controlled by varying the reaction time. Thin-film properties, including water uptake, swelling, and ionic conductivity, were characterized using ellipsometry, quartz crystal microbalance (QCM), Fourier transform infrared spectroscopy (FTIR), and electrochemical impedance spectroscopy (EIS).}

    \label{fig:experiment}
\end{figure}

To investigate the relationship between hydration and ion transport in polynorbornene-based anion exchange membranes (AEMs), we synthesized and functionalized thin films of poly(Bromo butyl norbornene) (PBBNB). PBBNB was synthesized via vinylic addition polymerization and spin-coated onto different substrates, chosen to accommodate various characterization techniques (see the extended methods section in the Supporting Information, SI appendix). The as-cast neutral films had an average dry thickness of approximately 60 nm, determined by ellipsometry. The films were then functionalized by trimethylamine (TMA) vapor infiltration, converting bromide groups to quaternary ammonium salts $\text{PBBNB}^{+}\text{Br}^{-}$. The ion exchange capacity (IEC) was controlled by varying the reaction times (2 h, 2.5 h, 5 h), yielding IEC values of 2.69, 3.05, and 3.43 mmol / g (see the SI Appendix, Fig. 2). The degree of functionalization was confirmed via FTIR spectroscopy (see the SI Appendix, Fig. 1). After functionalization the dry film thickness was measured to be approximately 100 nm for maximum functionalization.

The water uptake and swelling behavior were characterized using an \textit{in situ} RH generator-Ellipsometer-QCM system, which allows for the simultaneous measurement of film thickness and mass changes under controlled humidity. The system consists of an RH generator producing a precisely controlled humidified gas stream, an ellipsometer for real-time thickness monitoring, and a quartz crystal microbalance (QCM) for mass uptake measurements. A humidity sensor at the system outlet ensures precise RH control, and the data was acquired only after equilibrium was reached. Measurements of thickness expansion across different substrates and initial film thicknesses confirmed that the results obtained under the specific conditions required by each characterization technique were still comparable (see SI Appendix, Fig. 8).

The ionic conductivity, with bromide (Br\(^-\)) as the mobile charge carrier, was assessed via \textit{electrochemical impedance spectroscopy} (EIS) on interdigitated electrode arrays (IDEs) in a humidity-controlled chamber. The impedance spectra were fitted to an equivalent circuit model to extract the resistance of the film ($R_f$), which was then used to calculate the ionic conductivity (see the SI Appendix, Figs. 4 and 5).

This experimental framework enables a comprehensive analysis of how water content, swelling, and ionic transport are interrelated in thin-film AEMs. Full experimental details are provided in the Extended Methods section of the SI Appendix.

\subsection{Introduction to the Coupled Layers Model (CLM)}

\begin{figure*}[htbp]
    \centering
    \includegraphics[width=0.9\textwidth]{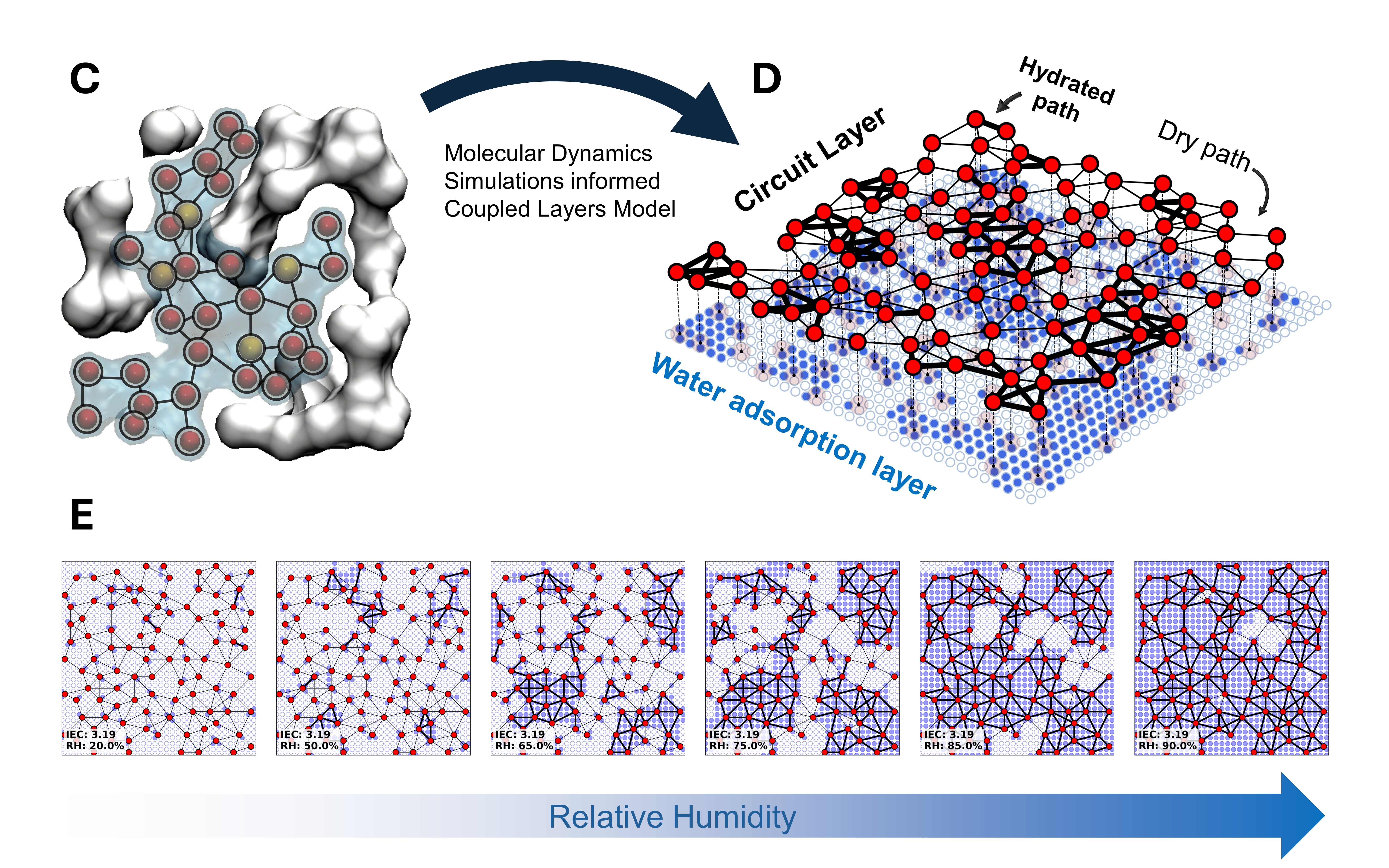}
    \caption{Description of the Coupled Layers Model. (A) A snapshot from previous MD simulations of the water and ion network that inspired the CLM model (image reproduced from Wang et al.~\cite{Zhongyang_Ge_AEM} under a Creative Commons Attribution 4.0 International License. The figure shows the polymer backbone as a white volume, water molecules in red, and bromide ions in yellow. (B) A typical simulation state of the model's two layers. In the first layer, red circles represent charged sites in the polymer matrix, and the edges connecting them represent the individual resistances of the circuit. Hydrated (fast) sites are depicted with thick lines, and dry (slow) sites with thin lines. The bottom layer corresponds to the water absorption lattice. Filled circles represent adsorbed water molecules, while empty circles indicate unoccupied potential sorption sites. Shaded regions highlight the sorption sites that belong to the ion's first hydration shell. (C) Typical equilibrium configurations produced by the CLM at increasing relative humidity from a top view. }

    \label{fig:simulation}
\end{figure*}

In this work, we introduce the Coupled Layers Model (CLM) to describe the interplay between water sorption and ionic conductivity in hydrated polymeric materials. The model consists of two interconnected but independently simulated layers: one representing water sorption thermodynamics and the other describing ionic transport through a resistor network. By explicitly coupling the two layers, the CLM captures the mutual dependence between water uptake and ionic conductivity, providing a simplified yet physically grounded framework to study polymeric membranes under different working conditions.

\paragraph*{\textbf{Creation of the Configurations:}}

In designing the CLM, one of our main goals was to preserve the essential features influencing the liquid-vapor phase behavior of water, namely the connectivity of the hydrogen bond network and the distance between hydrogen-bonded molecules. Additionally, we aimed to maintain the spacing between charged sites in the polymer in a manner consistent with previous MD studies. This ensured that the distance between charges matched the number of water molecules needed to "bridge" them, as observed in MD simulations. Finally, we selected the area for constructing the two layers to approach the experimentally observed number of water molecules per charge at maximum hydration.

With these objectives in mind, we created the network of water sorption sites by dividing a 10 nm × 10 nm area into a square lattice with a spacing of 0.3 nm, corresponding to the typical hydrogen bond distance. This resulted in 1089 sorption sites, which were connected using periodic boundary conditions to eliminate edge effects. This layer remained fixed throughout the simulations.

For the second layer of the model, we generated a random distribution of charge sites within the same area as the sorption lattice. This distribution represents the spatial arrangement of charged groups in the polymer matrix. The total number of charge sites in the network was determined by the degree of functionalization of the polymer. In our simulations, a network of 100 charge sites mapped well to an ion exchange capacity (IEC) of 3.05 mmol/g.

To ensure a physically reasonable charge distribution, we placed charges randomly while enforcing a minimum separation of 0.8 nm, inspired by the radial distribution function (RDF) observed between nitrogen groups in molecular dynamics simulations of similar polymers~\cite{Voth_simulations_AEM_overlap,simulations_allatoms-2022,simulations_water_shielding,simulations_bedrov_molinero_multiscale,simulations_bedrov_polarizable,Zhongyang_Ge_AEM}.  

The charges in the network were connected in the circuit if they were within 1.5 nm of each other. Although this arbitrary distance criterion impacted the overall conductivity of the material, it acted as a scaling factor for the conductivity curves without altering the relationship between conductivity and water content. For consistency, this distance was kept constant across all simulations. We averaged the simulation results from 250 initial random charge configurations for each tested condition.

Figure \ref{fig:simulation} (b) illustrates a typical simulation configuration, highlighting the two layers of the model and their interaction. The top layer represents the electric circuit formed by the random distribution of charges and their connections. Thicker lines indicate hydrated, low-resistance elements, while thinner lines indicate dry paths. The bottom layer represents water sorption, with blue circles indicating sites occupied by water molecules, and empty circles representing vacant sites. The shaded regions demarcate the water sites belonging to each charge's first hydration shell.

\paragraph*{\textbf{First Layer: Random Resistor Network (RRN):}}

In the first layer of the CLM, we represented the collective transport of ions in the polymeric matrix using an RRN solved numerically following Kirchhoff's laws~\cite{book_introduction_percolation}.

To construct the electric circuit, we treated the positions of the charges in the randomly generated network as nodes in a graph, with the edges connecting them representing independent resistors. The circuit was completed by linking all nodes on opposite borders of the graph to an external power source using connections of minimum resistance, ensuring a well-defined potential difference across the system. Based on Kirchhoff’s laws, the resulting system of equations was then solved to determine the current flow through the network, from which the total conductance was computed.

The magnitude of each individual resistance in the model was determined by the temperature and its hydration state. Hydrated connections between charges had lower ionic resistance compared to dry ones. We assumed the resistance between two charges to be independent of the distance separating them as long as they were connected and infinite if they were not. Furthermore, the temperature dependence of each resistance was modeled as independent activated processes according to the Arrhenius equation.

The system of linear equations derived from Kirchhoff's Current Law (KCL) and Kirchhoff's Voltage Law (KVL) was solved using the sparse matrix solver \texttt{linsolve.spsolve} from the SciPy library. The overall conductance of the system was converted into conductivity by considering the dimensions of the 2D lattice, assuming an arbitrary thickness of 1 nm. The conversion was carried out using the relation $\sigma = \frac{G L}{A}$, where $\sigma$ is the conductivity, $G$ is the overall conductance, $L$ is the characteristic length of the lattice, and $A$ is the cross-sectional area. 

The conductivity was calculated after each sampling step of the Monte Carlo simulation with the updated hydration states of each edge of the graph.

\paragraph*{\textbf{Second Layer: Monte Carlo Simulation of Water sorption}:}
We performed lattice Monte Carlo simulations in the grand canonical ensemble to study water sorption in the polymer. The Hamiltonian \( \mathcal{H} \) was given by:

\begin{multline}
\mathcal{H} = -E_{\text{ion}} \sum_{i \in \text{hydration}} n_i - E_{\text{HB}} \sum_{\langle i,j \rangle} n_i n_j \\
- E_{\text{pol}}(T) \sum_{i \in \text{lattice}} n_i - \mu \sum_{i \in \text{lattice}} n_i
\end{multline}

Where \( n_i \) is the occupancy state (0 or 1) of each site \( i \) of the lattice, \( E_{\text{ion}} \) represents the ion hydration energy, \( E_{\text{HB}} \) is the Water-Water interaction energy (hydrogen bond)  between neighboring water molecules, \( E_{\text{pol}} \) encapsulates all the polymer's contributions to water rejection (including polymer-polymer attraction, cation-anion attraction, and the work of expanding the polymer), and \( \mu \) is the chemical potential of the gas phase. Note that the ion-hydration energy affects only the four closest lattice sites to each charge in the first layer (RRN). Since the polymer is not explicitly represented in the simulations, we added a polymer water rejection energy as an effective potential term dependent on temperature as \( E_{\text{pol}}(T) = E_{\text{pol}}^{0} / (k_B T) \).

Here, the water-water interaction energy and the chemical potential of water vapor at various temperatures and relative humidities were optimized to approximate the critical point of real water as closely as possible. For detailed information on the finite size scaling analysis and the phase behavior of the water model, including curves of saturation pressures versus temperature, please refer to the SI Appendix, Figure 15. By fixing the critical temperature at 641.39 K, the model's critical pressure was determined to be 261.7 atm, which compares reasonably well with the actual value of 218 atm. With this parametrization, the water-water interaction energy was set at -8.28 kJ/mol. The remainder of the interaction parameters were determined to approximate experimental observations. 

The simulations began by initializing the lattice in an empty state at the lowest relative humidity (RH). For each subsequent RH level, the final configuration from the previous RH step was used as the new starting configuration. The energy of each lattice site was determined from the Hamiltonian on the basis of its current occupancy state.

During each Monte Carlo step, a site was randomly selected and its state was flipped. The flips were always accepted if the change in energy (\(\Delta E\)) was \(\leq 0\); if \(\Delta E > 0\), they were accepted with a probability of \(e^{-\Delta E / kT}\).

This process was repeated for \(1500 \times N\) steps (where \(N\) is the total number of sites) to reach chemical equilibrium. After equilibration, additional \(100 \times N\) steps were performed to generate equilibrium configurations for analysis. During this procedure, every site was sampled with equal probability, ensuring adherence to the principle of detailed balance.

\paragraph*{\textbf{Coupling Between Layers:}}
Central to the CLM is the coupling between the Random Resistor Network and the water sorption layers. The charges in the first layer determine the preferential water sorption sites in the second layer. Conversely, the water absorbed in the path connecting two charges in the second layer determines the resistance of that connection in the first layer.

Specifically, the four water sites closest to each charge in the RRN layer were considered to be part of the ion’s first hydration shell and have a very favorable sorption energy. Water sites within half a water-water distance (0.15 nm) to the line connecting two charges in the RRN defined the hydration state of that path. A path was defined as ``hydrated'' if all associated water sites are occupied and ``dry'' otherwise.

The Hamiltonian of the Monte Carlo simulation was updated with the positions of the ion hydration sites with each new random distribution of charge. In turn, the hydration state of the edges of the RRN was updated after each equilibrium sampling step of the Monte Carlo simulation.

\paragraph*{\textbf{Model's Parametrization:}}

All model parameters were determined using a polymer with a degree of functionalization of 85\% (IEC = 3.05 mmol/g) at 300 K as a reference state. The parametrization was based on experimental data at 20\% and 85\% relative humidity (RH), which were the only two calibration points used. Rather than performing a formal optimization, parameters were selected through trial and error by running test simulations on a single network topology.  

For activated processes that govern ionic conduction, the total apparent activation energies at 20\% RH and 85\% RH were extracted from the experimental data and rounded to 40 kJ/mol and 20 kJ/mol, respectively. To complete the parametrization, the preexponential factors were adjusted to match the experimental conductivity values at 300 K for both reference cases. Under this parametrization, the individual conductances were defined as functions of temperature:  

\begin{equation}
\sigma_{\text{hydrated}} = 1.38 \times 10^{-6} \, \text{s} \cdot e^{\left(-\frac{20 \, \text{kJ/mol}}{k_B \cdot T}\right)}
\end{equation}

\begin{equation}
\sigma_{\text{dry}} = 1.25 \times 10^{-7} \, \text{s} \cdot e^{\left(-\frac{40 \, \text{kJ/mol}}{k_B \cdot T}\right)}
\end{equation}

Using the same reference points at 20\% RH and 85\% RH, we varied the interaction parameters in the Monte Carlo simulation to approximate the experimentally observed water uptake. The parameter set that best reproduced the experimental data was: \(E_{\text{HB}} = -8.28\,\text{kJ/mol}\), 
\(E_{\text{ion}} = -20\,\text{kJ/mol}\), 
\(E_{\text{pol}}^0 = 5.75\,(\text{kJ/mol})^2\). 

\section{Results}

\subsection{Water Uptake, Conductivity, and Volume Expansion Ratio}

\begin{figure*}[htbp]
    \centering
    \includegraphics[width=0.9\textwidth]{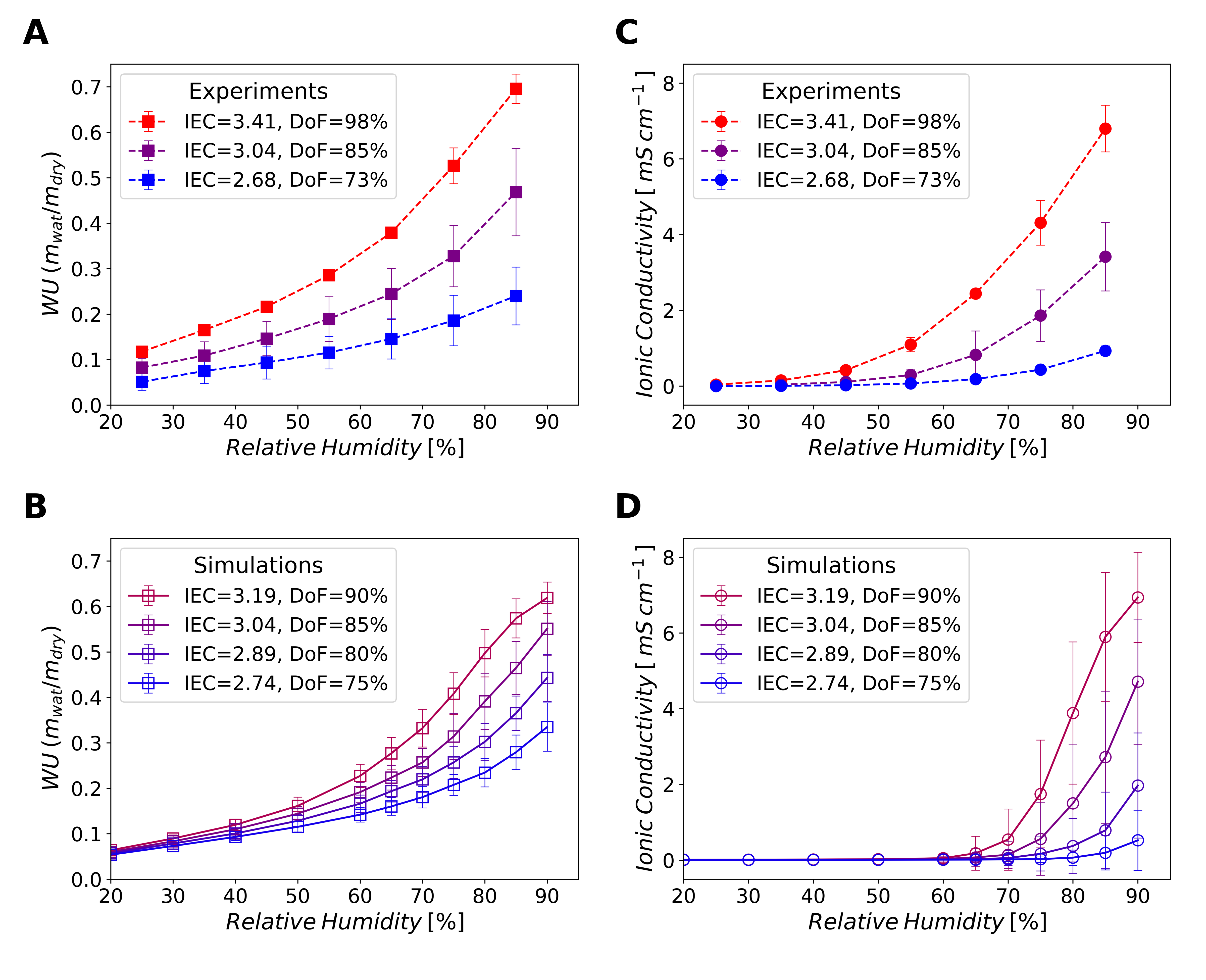}
    \caption{Membrane's performance characterization. Water uptake and as a function of relative humidity for various IECs: (A) experiments; (B) simulations. Ionic conductivity as a function of relative humidity for various IECs: (C) experiments; (D) simulations.}
    \label{fig:water_content}
\end{figure*}

Water uptake (WU) is a dimensionless value defined as the mass of water absorbed (\(M_{\text{water}}\)) divided by the mass of the dry PBBNB\(^+\)Br\(^-\) thin film (\(M_{\text{dry}}\)). The methods for measuring WU in thin films include quartz crystal microbalance (QCM) and density approximation. Detailed procedures are explained in the Extended Experimental Methods Section in the SI Appendix.

\begin{equation}
WU=\frac{M_{water}}{M_{dry}}
\end{equation}

The ionic conductivity (\(\sigma\)) was extracted from electrochemical impedance spectroscopy (EIS) measurements performed on top of interdigitated electrode arrays (IDE) using the following equation:

\begin{equation}
\sigma=\frac{1}{R_f}\frac{d}{l\left(N-1\right)h}
\end{equation}

where \(R_f\) is the ionic resistance, \(d = 8\ \mu\text{m}\) is the distance between the adjacent electrode teeth, \(l = 500\ \mu\text{m}\) is the effective electrode length, \(N = 80\) is the number of electrode teeth and \(h\) is the thickness of the film.

Figure \ref{fig:water_content} (a to d) presents results of water absorption and conductivity collected experimentally (top row) and from simulation (bottom row) at room temperature. In the figure, the color range goes from blue for polymers with low functionalization to red for high functionalization. In Figures \ref{fig:water_content} (a) and (b), the water uptake initially increases gradually with increasing relative humidity for all polymers regardless of their IEC. However, as relative humidity continues to increase, water uptake enters an exponential regime and differences emerge between polymers with different degrees of functionalization. In this regime, polymers with higher IEC absorbed significantly more water than polymers with lower IEC.

The right column of the figure shows the ionic conductivity obtained experimentally from EIS (c) and calculated from simulations (d). Similarly to water absorption, the conductivity remains very low at low RH and rapidly increases at higher RH. In both low- and high-conductivity regimes, thin films with a higher IEC exhibit significantly higher ionic conductivity at the same RH. This figure highlights two key points: first, polymers with lower IEC appear to transition to the high-conductivity regime at higher RH values compared to polymers with higher IEC; second, in the computational model, these transitions are shifted to higher RH values, probably due to the reduction in dimensionality of the simulation from 3D to 2D.

During the measurements, the polymers swelled significantly as the water content in the films increased. Our experiments showed that volume expansion was directly proportional to the water content, suggesting minimal excess volume from absorption. Plots of thickness expansion as a function of water uptake, along with a brief discussion on the impact of swelling on conductivity and its implications for our model, can be found in the SI appendix.

\subsection{Ionic Conductivity as a Function of Water Content}

\begin{figure*}[htbp]
    \centering
    \includegraphics[width=0.9\textwidth]{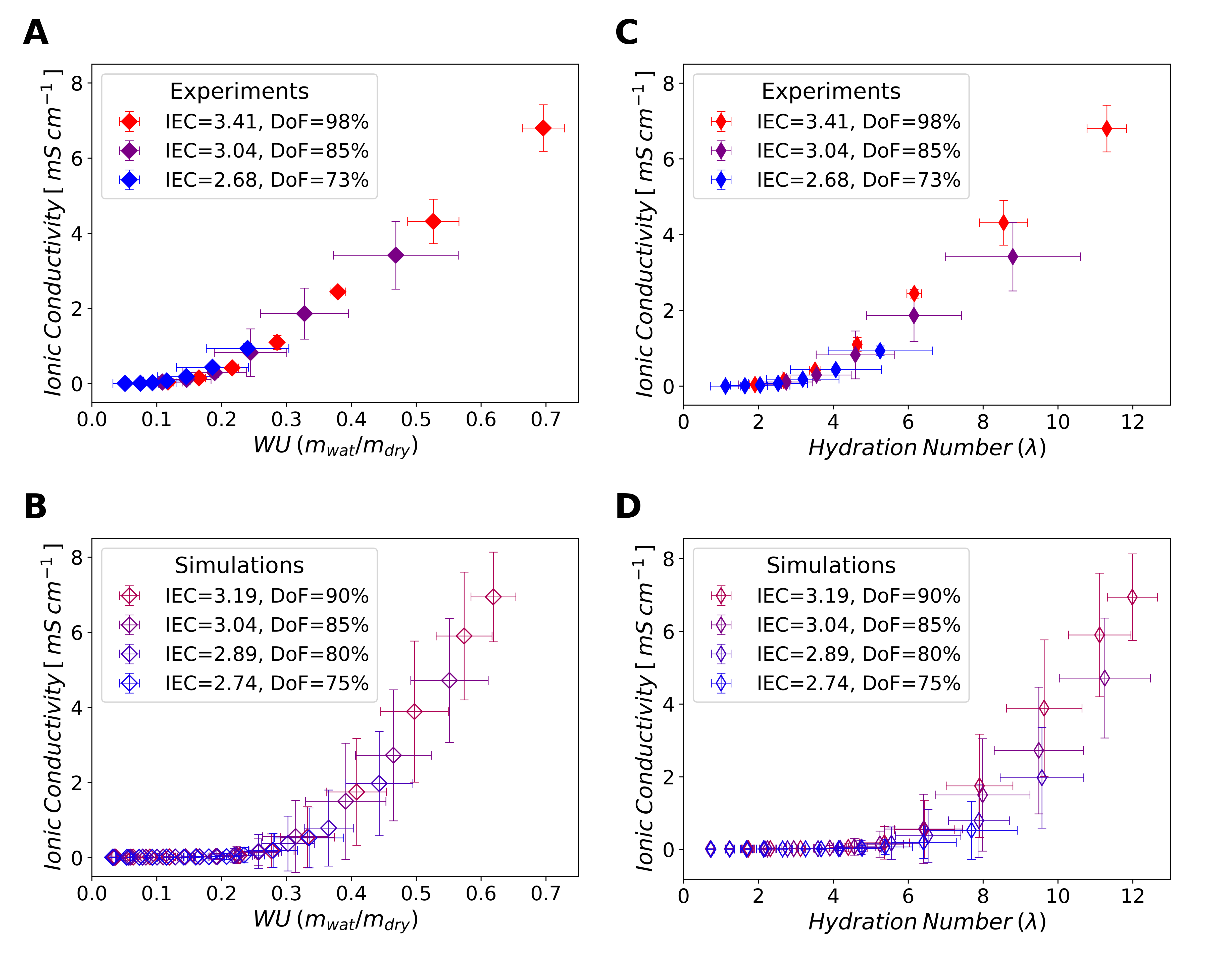}
    \caption{Universal ionic conductivity as a function of water content. Figures (A) and (B) show how the conductivity curves for polymers with different IECs converge onto a single master curve when plotted against water uptake. Figures (C) and (D) present the same conductivity data plotted against the hydration number, where the curves do not collapse. The top row of the figure shows data obtained from experiments, and the bottom row from simulations.}
    \label{fig:mastercurve}
\end{figure*}

Figures~\ref{fig:mastercurve} (a) and \ref{fig:mastercurve} (b) show the relationship between ionic conductivity and water uptake for all polymers tested in experiments (top) and simulations (bottom). Our primary observation was that all of the ionic conductivities aligned along a master curve when plotted against the water content. Polymers with a higher IEC extended further along this curve, achieving greater water uptake and ionic conductivity. Furthermore, Figure~\ref{fig:mastercurve} underscores the success of CLM in capturing the relationship between conductivity and water content observed in experiments.

Panels~\ref{fig:mastercurve}(c) and \ref{fig:mastercurve}(d) illustrate the relationship between conductivity and the hydration number (\(\lambda\)). The hydration number represents the number of water molecules per ionic group. The relationship between \(\lambda\) and WU can be expressed as follows:

\begin{equation}
\lambda=\frac{1000}{18.02}\frac{WU}{IEC}
\end{equation}

Unlike in Figures \ref{fig:mastercurve}(a) and (d), the curves in Figures \ref{fig:mastercurve}(c) and (d) did not collapse into a single curve. Polymers with lower IEC exhibited lower conductivities for any given \(\lambda\) in experiments and simulations. This result is noteworthy because other studies have reported similar universal conductivity behavior when plotted against \(\lambda\), particularly for Nafion\textregistered{} and other proton exchange membranes.

Finally, we observed a similar universal behavior to that shown in Figures \ref{fig:mastercurve}(a) and (b) when using the water volume fraction as a descriptor of the hydration state of the films (see the SI Appendix, Fig. 9). Although the water volume fraction is commonly used to describe the water content in models such as the percolation theory~\cite{kirkpatrick1973percolation} and the Mackie and Meares model~\cite{classic_flux_equations}, we chose to use water uptake (WU) in the main text to simplify the analogy between the 3D polymer network and the 2D simulation model. However, as shown in the SI Appendix, water uptake, water volume fraction, and water concentration can be used interchangeably, as their respective curves collapse into a single master curve for all polymers tested.

\subsection{Water Sorption and Local Charge Density}
\begin{figure*}[htbp]
    \centering
    \includegraphics[width=0.9\textwidth]{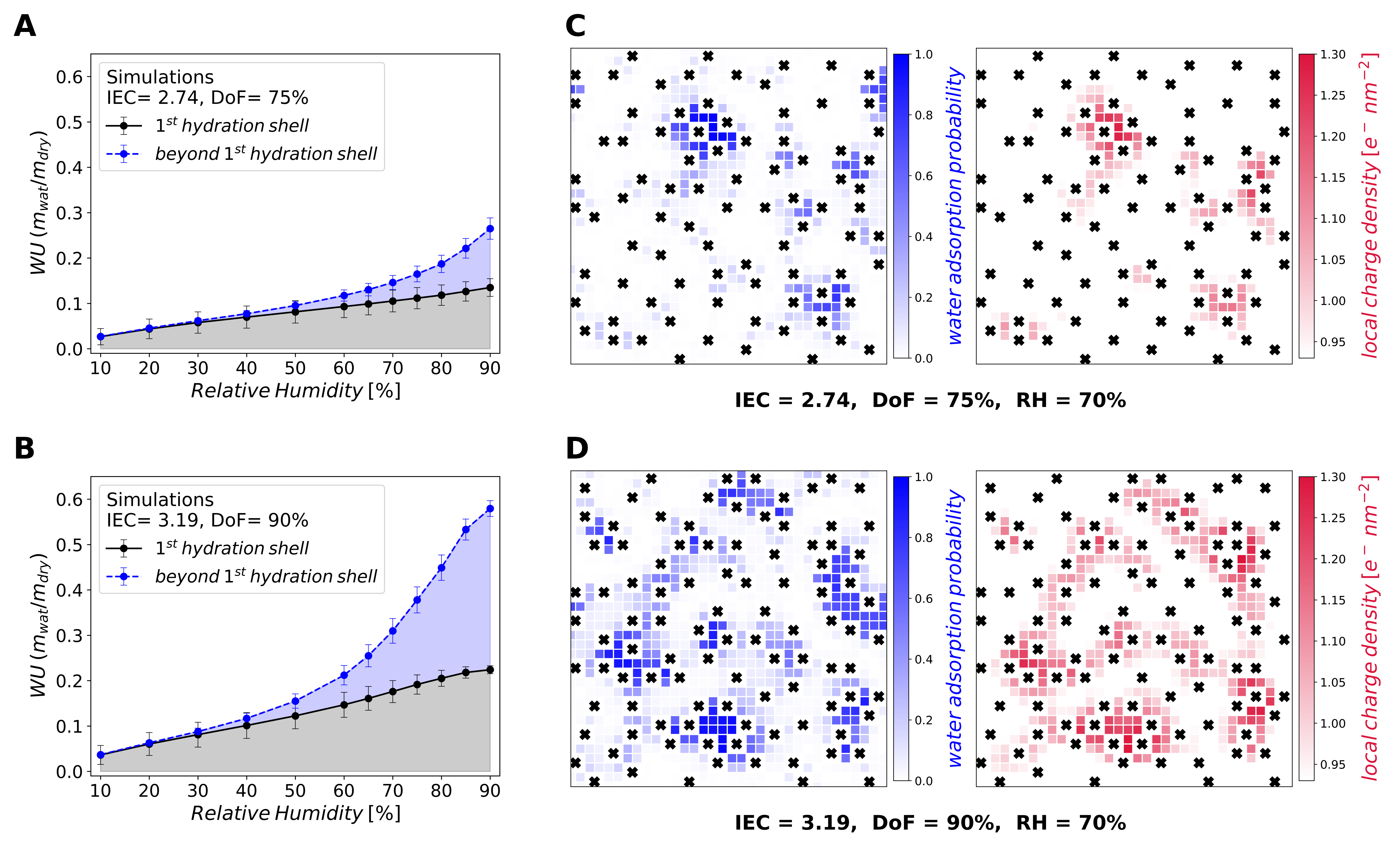}
    \caption{Water absorption and local charge density. The left column shows water sorption isotherms at room temperature for polymers with low IEC (A) and high IEC (B). The isotherms are divided into water adsorbed in the ion's first hydration shell (black) and beyond the first hydration shell (blue). Figures (C) and (D) compare the water sorption probability at each individual site of the lattice (left) with the local density of charges at each site (right). The top row corresponds to a polymer with low IEC, while the bottom row corresponds to a polymer with high IEC. Crosses indicate the positions of polymer charges in the simulation. Lattice sites that belong to the first hydration shell are painted white, regardless of their occupation state, to emphasize absorption beyond the first shell. In these figures, the relative humidity (RH) was 70\% for both polymers. We chose 70\% RH because higher RH saturates the sorption probability at 1, making the comparison more difficult }
    \label{fig:charge_density}
\end{figure*}

In this section, we analyze the water sorption process in relation to the charge density within the polymer matrix. This analysis is relevant because much of the literature focuses on the importance of various ``thermodynamic'' types of water for the ion conduction mechanism. Common terms used to describe different types of water include free versus bound water, freezable and non-freezable water, etc.~\cite{original_freezable_water,freezable_water_recent,two_types_of_water_clusters}. Beyond semantics, the correlation between changes in water dynamics and thermodynamics and the onset of increased conductivity has been verified. Specifically, in what is commonly termed the theory of water clustering in polymers, the presence of free water is considered necessary for vehicular ion conduction~\cite{}. Furthermore, previous studies have identified the presence of water beyond the first hydration shell of ions as a strong indicator of the onset of enhanced ionic conduction~\cite{Zhongyang_Ge_AEM,Voth_simulations_AEM_overlap,local_hydration,QENS_water_ions_dynamics,simulations_water_shielding,weaker_electros_water_MM}.

In Figure \ref{fig:charge_density}, we first present water absorption isotherms at room temperature for polyelectrolytes with IEC values of 2.74 mmol/g and 3.19 mmol/g. In our simulations, only the four nearest water molecules interact with the ions and thus are considered part of the ion's first hydration shell. Beyond the nearest neighbors of the ion, water can only interact with other water molecules. However, it is important to note that the thermodynamic stability of water beyond the first hydration shell is a cooperative process that strongly depends on the presence of adjacent water molecules~\cite{Lattice_adsorption_aranovich_2}. For example, a second-shell hydration site surrounded by four first-shell sites is much more likely to capture water than a site adjacent to only one first-shell site. 

From the comparison between Figures \ref{fig:charge_density} (a) and (b), two key points emerge. First, the exponential increase in water uptake begins when absorption extends beyond the ion's first hydration shell. Second, absorption beyond the first hydration shell varies significantly between polymers with different IECs. In contrast, water absorption within the ion's first hydration shell (shaded gray in the figures) would be approximately the same for both polymers if expressed as a hydration number.

To further evidence the cooperativity of absorption beyond the first hydration shell, Figures \ref{fig:charge_density} (c) and (d) compare the probability of sorption of each individual hydration site in the network (left) with the local charge density of the same sites (right). 

We calculated a smoothed local charge density using Gaussian kernel density estimate (see methods) with a bandwidth of 0.2 nm. The local probability density was normalized by the total number of charges in the polymer to obtain the final density. With these smoothing parameters, the influence of the charges extends over a range equivalent to two to three water hydration shells at T = 300K.

In the plots, the positions of the charges are indicated with black crosses. For clarity, all first-shell hydration sites are rendered white to avoid interfering with the visual identification of the occupancy state of the second and third hydration shells. The probability of water sorption in each site beyond the first hydration shell is rendered in shades of blue in the left panel, while the local charge density of the same sites is shaded red in the right panel.   

The figure illustrates that, at a given relative humidity, regions beyond the ion's first hydration shell with higher local charge density exhibit a significantly higher probability of water sorption compared to regions with lower local charge density~\cite{two_types_of_water_clusters,adsorbtion_model_theory_1}. Although polymers with higher overall charge density (IEC) naturally absorb water more efficiently~\cite{collapse_volume_fraction_MM}, it is essential to note that the relationship between local charge density and water sorption probability remains the same in all polymers examined. 

An interesting observation from our sorption studies in the simulations was the presence of hysteresis (see the SI Appendix Fig. 14), which was directly related to the presence of water beyond the first hydration shell. This observation agrees well with previous studies of adsorption in narrow pores~\cite{monson2001Lattice_condensation,monson2012hysteresis,evans1986_Capilary_condensation_hysteresis,evans1990_Capilary_condensation_hysteresis}. Recently, Chen et al. reported a similar finding attributable to the characteristics of the hydrogen bond network of water ~\cite{hysteresis_Carmeliet_Hydrogen_Bond}. 

These findings highlight that cooperative effects among water molecules significantly enhance water sorption beyond the ion’s first hydration shell. At higher local charge densities, increased interactions among water molecules strongly promote water uptake in regions between charges. In contrast, within the first hydration shell, water uptake is dominated by strong ion–water interactions, making it relatively insensitive to cooperative effects.

\subsection{Conductivity Scaling With Water Content and Percolation}

\begin{figure*}[htbp]
    \centering
    \includegraphics[width=0.9\textwidth]{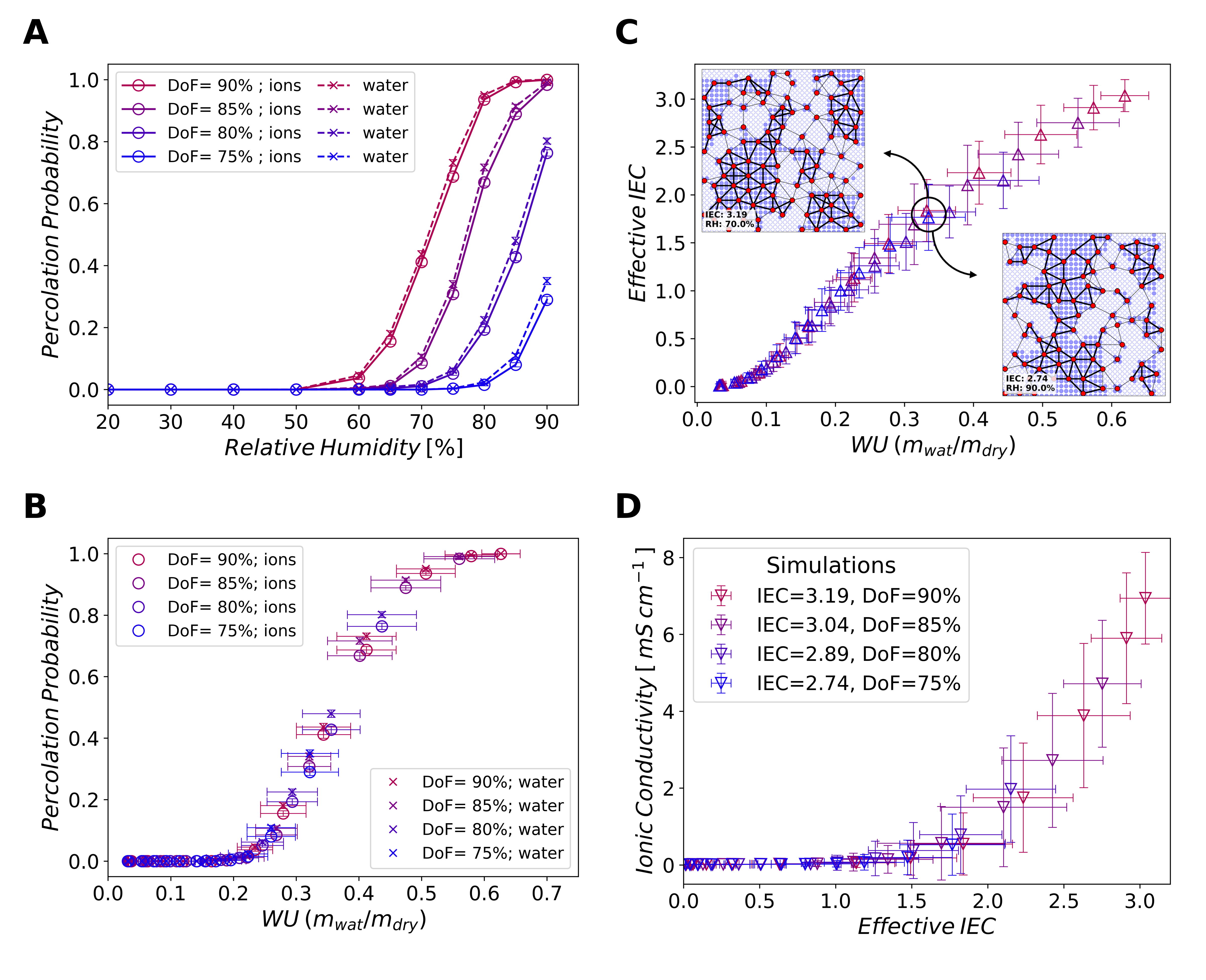}
    \caption{Joint percolation and effective IEC as the reasons for the conductivity universal scaling with water content. (A) Percolation probability of water and ions networks showing simultaneous percolation. (B) Percolation probability, when plotted against water content, exhibits the same universal scaling observed for conductivity. (C) Universal relationship between water content (as WU) and the number of ionic groups that are hydrated and effectively contributing to ion transport (Effective IEC). The insets show representative configurations of polymers with high and low actual IEC at different RH, but with similar water content and Effective IEC. (D) Conductivity as a function of Effective IEC, showing the same universal scaling noted with water content.}
    \label{fig:conductivity_percolation}
\end{figure*}

A central question in IEM research is why the ionic conductivity is so strongly related to the water content of the polymer film~\cite{water_isotherms_dekel,water_degrades_mec,conductivity_vs_lambda_volume_fraction,original_freezable_water,freezable_water_recent}. The percolation analysis shown in Figure \ref{fig:conductivity_percolation} aims to establish causality in the correlation between conductivity and water content observed in Figure \ref{fig:mastercurve}.

The first important result of this section, shown in Figure \ref{fig:conductivity_percolation}(a), is that in all polymers the percolation probability of the ion network is nearly identical to the percolation probability of the water hydrogen bond network.

To better understand Figure \ref{fig:conductivity_percolation},  it is important to recall that in our model, a pathway between two charges is considered ``fast'' only if all adjacent sorption sites along that pathway are occupied by water. This definition makes the percolation of fast ionic sites contingent on the formation of a continuous water network. However, if this were the only condition, one would expect the percolation of a hydrated ion network to occur only after the water network itself has already percolated. In other words, the emergence of ionic connectivity would typically lag behind water percolation.

In contrast, our simulations reveal that whenever a continuous water network percolates through the system, a percolating network of fast ionic sites emerges simultaneously. This surprising result arises naturally from the preferential formation of water bridges between charges in regions of sufficiently high local charge density, as described in the previous section. In contrast, regions with low charge density exhibit negligible water sorption, which prevents water from percolating independently.

In summary, Figure \ref{fig:conductivity_percolation} (a) shows that the percolation of hydration water and the percolation of fast ionic pathways are closely coupled in these systems, and neither occurs independently.

In Figure \ref{fig:conductivity_percolation} (b), we further demonstrate that the probability of percolation, like the conductivity, follows a universal trend across all systems when expressed as a function of the water content.

This result significantly simplifies the understanding of the percolation process. Predicting the percolation of the polymer’s charge network is complex, requiring specific definitions such as the network’s connectivity and the volume of its elements, which are unique to each polymer. In contrast, predicting the percolation of the water network is much simpler because its connectivity is determined by the hydrogen bond network topology, which is likely independent of the material. Additionally, percolation theory indicates that the percolation probability directly depends on the volumetric proportion of the element forming the network~\cite{book_introduction_percolation}. The percolation threshold in our simulations was determined from Figure \ref{fig:conductivity_percolation}(b) to be around 0.35 in WU (or 0.6 in filling fraction) in the lattice, which perfectly matches the expectation for a square lattice~\cite{mertens2022percolation_treshold_square}. For a tetrahedral lattice in 3D, the percolation threshold occurs at a volume fraction (or filling fraction) of around 0.147~\cite{frisch1961percolation_treshold_ice}. While we cannot precisely estimate the water volume fraction in the experiments or determine the exact percolation threshold, the plot of conductivity versus volume fraction in the SI Appendix, Figure 10, shows that conductivity measured experimentally begins to increase around this point, much like the increase observed in simulations near the percolation threshold.

In addition to the percolation analysis, we introduce the concept of effective IEC to better illustrate how water influences ion transport in polymer films. Unlike the total IEC, which accounts for all ionic groups present in the polymer, the effective IEC includes only the subset of ionic groups that are hydrated and actively contribute to ion conduction.

To calculate the effective IEC in our simulations, we first determine the effective charge concentration, which represents the density of hydrated ions. This is defined as the total number of hydrated ion paths, shown as thick black bars in the insets of Figure~\ref{fig:conductivity_percolation}(c), divided by the mean degree of the graph, which is the average number of connections each charge has with its neighbors in the conduction network. We denote this as $\rho_\text{eff} = N_\text{hydrated} / \langle k \rangle$. The effective ion exchange capacity is then estimated as $\text{IEC}_\text{eff} = \text{IEC}_\text{total} \times (\rho_\text{eff} / \rho_\text{total})$, where $\text{IEC}_\text{total}$ is the nominal ion exchange capacity and $\rho_\text{total}$ is the total charge concentration.

Figure \ref{fig:conductivity_percolation} (c) shows the dependence of the Effective IEC on water content. This figure reinforces what was previously shown in Figure \ref{fig:charge_density} about the conserved ratio between absorbed water and the number of hydrated charges. The insets in the figure display two typical configurations of high and low IEC polymers at different relative humidities (RH) but with approximately the same water content and effective IEC. Interestingly, the patterns of water absorption and the formation of hydrated paths connecting charges are very similar in both cases, which is consistent with previous simulations of polymers with different IECs~\cite{molinero_isotherms}. Panel (d) of Figure \ref{fig:conductivity_percolation} complements this analysis by showing that the effective IEC also converges all the conductivity curves into one. This indicates that the water content determines the total number of charges that participate effectively in conduction out of the total number of charges in the polymer.

\subsection{Temperature Effect and Apparent Activation Energy}

\begin{figure*}[htbp]
    \centering
    \includegraphics[width=0.9\textwidth]{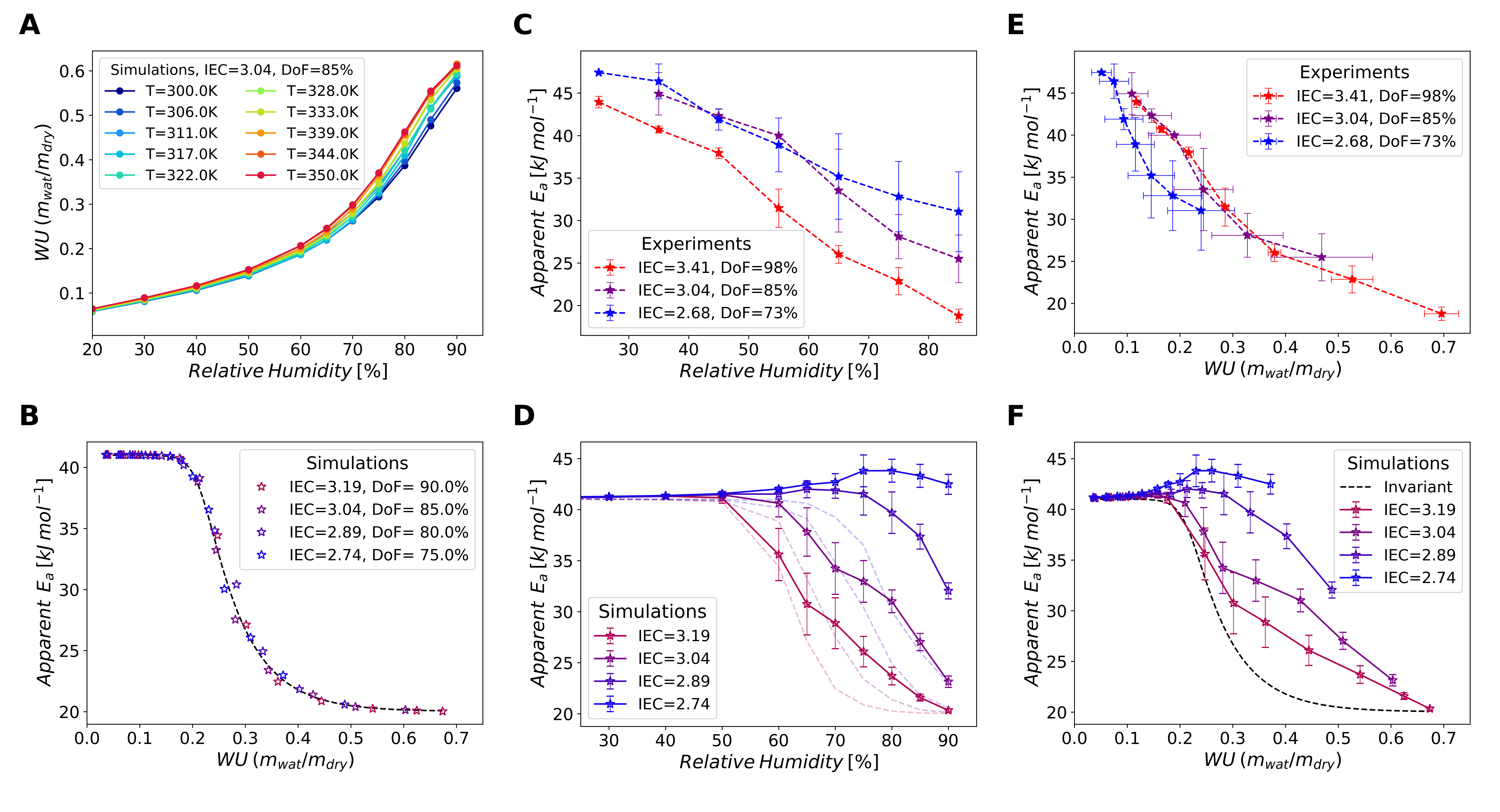}
    \caption{Effect of temperature on membrane performance. (A) Variations in water sorption isotherms from simulations for the polymer with 3.04 mmol/g IEC. (B) Activation energy curves versus water uptake from simulations with a fixed absorption isotherm at 300K. (C) Experimentally measured apparent activation energies versus RH. (D) Apparent activation energies from simulations, with faded dashed lines indicating the expected activation energy if water uptake was insensitive to temperature changes. (E) Activation energy curves from (C) plotted against water uptake instead of RH. (F) Apparent activation energies from (D) plotted against water uptake instead of RH. The dashed line in (F) corresponds to the behavior shown in (B), evidencing the deviation of the apparent activation energy from a pure Arrhenius process where water absorption does not vary with temperature.}
    \label{fig:temperature_impact}
\end{figure*}

To conclude our presentation of results, we consider how temperature affects the ionic conductivity in the polymer film. This analysis is particularly challenging because conductivity is highly sensitive to various factors, requiring precise control of all variables. Studying temperature effects is important because it is often used to probe the conduction mechanism, though the effect of temperature is typically interpreted under the assumption of constant water content in the polymer. To the best of our knowledge, this is the first simulation to simultaneously analyze the effects of temperature on both ionic conductivity and water absorption.

Temperature can influence ionic conduction in two ways. Firstly, through changes in the probability of overcoming the activation barriers that retain the ions in place. This factor is incorporated into our model as an independent thermally activated process for both hydrated and dehydrated individual resistances. Secondly, temperature affects conductivity by modifying the water absorption process. At a fixed RH, temperature impacts the simulations in multiple ways: it modifies the acceptance probability of energetically unfavorable steps via the Metropolis criterion, slightly shifts the chemical potential of water vapor (see SI Appendix, Fig. 15) and alters the effective water rejection potential.

\begin{figure*}[htbp]
    \centering
    \includegraphics[width=\textwidth]{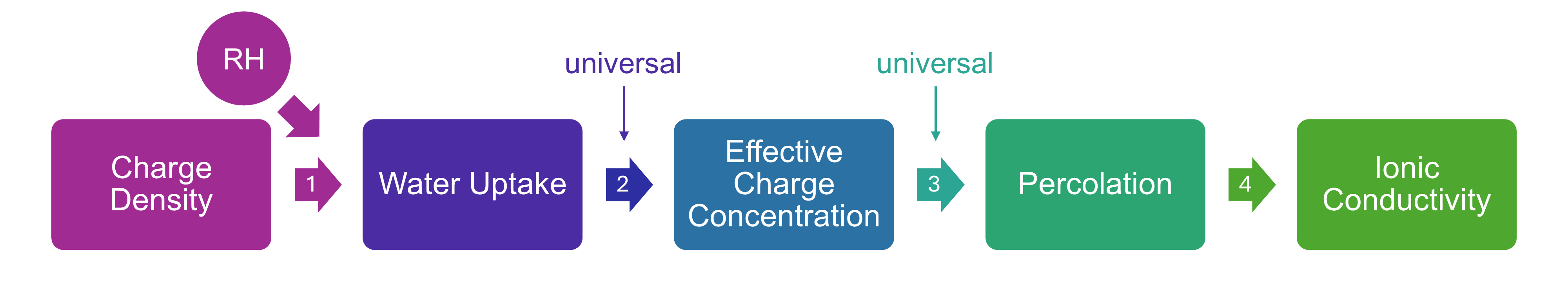}
    \caption{Causal sequence linking ion exchange capacity (IEC) to ionic conductivity. At a given relative humidity (RH), polymers with higher IEC absorb more water (1). However, the number of hydrated pathways connecting charges (effective IEC) per unit of absorbed water is nearly identical across all polymers (2). The effective IEC then determines the probability of forming a percolating network of hydrated domains (3), which in turn governs the macroscopic ionic conductivity (4). Steps 2 and 3 unify the behavior observed across all tested polymers.}
    \label{fig:mechanism}
\end{figure*}

Figure \ref{fig:temperature_impact} (a) displays the sorption isotherms for each temperature obtained from simulations at 3.04 IEC. Although the isotherms are similar, water uptake varies by nearly 15\% between the highest and lowest temperatures at high relative humidity. This variation is particularly significant because, as shown in Figure \ref{fig:mastercurve}, conductivity is highly sensitive to the water content in this regime. Even small changes in the WU can lead to large changes in the conductivity. Figure \ref{fig:temperature_impact}(b) highlights the impact of sorption changes on ionic conductivity by displaying activation energy curves from simulations in which the sorption isotherm was deliberately kept constant despite temperature variations. In these simulations, the water content was arbitrarily fixed at 300K, and only individual resistances were allowed to change with temperature. As a result of keeping the isotherm fixed, all curves perfectly align when plotted against the water content. Three distinct regimes emerge: at low RH, activation energy is dominated by the dehydrated components; at high RH, it is dominated by hydrated components; and between them, there is a quick transition between the two regimes. 

Since temperature affects not only the mobility of the ions but also the sorption of water, we refer to the measured slope of the conductivity versus the inverse temperature as ``apparent'' activation energy, which is a common practice in processes that involve coupling between successive activated processes or include temperature effects on chemical equilibrium constants. 
It is worth noting that despite the complex relationships mentioned above, the ionic conductivity in PBBNB\(^+\)Br\(^-\) thin films,  (\(\sigma\)) always exhibited Arrhenius-like behavior in our experiments and simulations, evidenced by the linear relationship:

\begin{equation}
\ln(\sigma) = \ln(\sigma_e) - \frac{E_a}{RT} 
\end{equation}

where \(\sigma_e\) is the preexponential factor, \(R\) is the gas constant (8.314 J mol\(^{-1}\)K\(^{-1}\)), and \(T\) is the absolute temperature (in Kelvin). \(E_a\) is the activation energy. Please refer to the SI Appendix Figures 11 and 12 for plots of $\ln(\sigma)$ versus $T^{-1}$ obtained in experiments and simulations.

Figure \ref{fig:temperature_impact} also shows the apparent activation energies as a function of relative humidity, obtained experimentally (c) and from simulations (d). The global apparent activation energies were determined from the slope of the plot of \(\ln \sigma\) versus \(\frac{1}{T}\) for each given RH. 

Our experimental results, consistent with previous reports, revealed a change in the slope of the apparent activation energy curve that was especially notable for polymers with high IEC. The slope became steeper when the conditions transitioned from low to high RH. The observed change in slope has been linked to a shift in the transport mechanism, which occurs as water begins to absorb into the second hydration shell of the ions~\cite{Zhongyang_Ge_AEM}. That study also demonstrated that the slope transition correlates not only with a shift in ionic conduction but also with changes in the rotational relaxation dynamics of hydration water. In our experiments, the RH at which this transition occurred shifted to a higher RH as the polymer's IEC decreased, eventually disappearing at the lowest IEC.

In the CLM simulation, we observed a similar change in the slope of the curves following a plateau at low relative humidity. In addition, some activation energy plots show a positive slope and a peak at medium to high relative humidity levels. To highlight the impact of changing water content, we include as dashed lines the activation energy curves corresponding to the fixed water content case presented in panel b.

The third column of Figure \ref{fig:temperature_impact} presents the same results as in (c) and (d), but plotted against water uptake instead of RH. These figures reveal that the curves for different polymers do not collapse as they did for other variables at the same temperature. However, the degree of overlap was better in the experiments than in the simulations, suggesting perhaps a lower variation in the experimental sorption isotherms. The similarities between the experimental and simulation results are clear, although the experimental curves appear to be shifted to higher RH values, missing the initial plateau region. A similar change in experimental curves was observed in the data of conductivity versus water uptake shown in Figure \ref{fig:water_content}, and was attributed to differences in dimensionality between the experiments and the 2D simulations.

\section{Discussion}

One of the key findings of this work is the seemingly universal relationship between the water content in membranes and the ionic conductivity, a phenomenon frequently reported in the literature, yet not clearly understood. Based on our analysis, we find that ionic conduction in the membrane proceeds through a causal sequence, illustrated in Figure \ref{fig:mechanism}.

Figures \ref{fig:water_content} and \ref{fig:charge_density} show that water uptake is controlled by local charge density in combination with relative humidity. We showed that polymers with higher IEC exhibit more regions where local charge accumulation is sufficient to retain water. Consequently, polymers with higher IEC absorb more water under identical RH conditions (step 1 in Figure \ref{fig:mechanism}). 
However, we also showed in Figures \ref{fig:charge_density} and  \ref{fig:conductivity_percolation} that the ratio between the number of hydrated charges and the amount of water absorbed is constant across all polymers regardless of their IEC. In other words, the water content alone determines what we call the "effective IEC", which is precisely the number of ions engaged in fast hydrated conductive paths (step 2 in Figure \ref{fig:mechanism}). This relationship explains why water uptake, or equivalently, the water volume fraction, serves as an effective normalizing factor of membrane behavior, whereas the hydration number ($\lambda$) does not.

The next factor underlying the universal conductivity curves shown in Figure \ref{fig:mastercurve} is the relationship between the effective IEC (or water content) and the formation of a continuum path of fast conduction (step 3 in Figure \ref{fig:mechanism}). Our analysis shows that the probability of percolation depends solely on the water content and is independent of the IEC of the polymer, indicating that the underlying topology of the percolating network is conserved between polymers. For the Coupled Layers Model, we demonstrated that this network is defined by the topology of the hydrogen-bond network of water. While the data clearly support this mechanism for the CLM, the experimental results shown in SI Appendix, Fig. 10, also provide some preliminary evidence that a similar effect may also govern percolation in real polymer films.

Step 4 in Figure \ref{fig:mechanism} completes the connection between IEC, water uptake, and ionic conductivity by relating the percolation of the hydrated pathways to the macroscopic conductivity of the polymer. Although the role of cooperativity in ionic conduction was already recognized in the 1970s, it has been sometimes overlooked in some recent mechanistic studies of ion conduction in IEMs. Percolation is necessary because ion diffusion is a collective process controlled by charge conservation; in such a process, ionic current preferentially flows through regions of lower electrical resistance. Provided that local hydration substantially reduces the energy barriers for individual ion motion, the relative abundance and connectivity of hydrated and dry regions governs the overall conductivity.

At low water content, conductivity remains low and is dominated by the high resistance of dry regions. Even the presence of minority hydrated pathways has little impact until a single continuous hydrated path emerges, at which point the conductivity rises sharply. Beyond this percolation threshold, conduction is dominated by hydrated pathways, and any remaining dry domains do not significantly affect conductivity. However, increasing hydration further strengthens the robustness of the percolating network, causing the conductivity to continue to rise after the threshold is reached.~\cite{Zhongyang_Ge_AEM}

In summary, while our study offers clear insight into the behavior of the specific polymers we tested, further research is needed to determine whether the relationship between conductivity and water content applies more broadly. Several other universal relationships have been reported, such as between conductivity and hydration number (\(\lambda\)) ~\cite{conductivity_lambda_ions_noQ,kusoglu_weber_review_PEM,conductivity_vs_lambda_empirical}, conductivity and \(\lambda\)/ion charge~\cite{kusoglu_weber_ions_1,kosoglu_ions_1,weber_kusoglu_3D_network}, mobility and \(\lambda\)~\cite{Winey_experiments_and_simulations_1}, and conductivity and volume fraction~\cite{collapse_volume_fraction_MM,conductivity_vs_lambda_volume_fraction,kosoglu_ions_1}. Reconciling these relationships with our findings remains an important open question. Moreover, the water–conductivity relationship identified here may not extend to polymers with different chemistries, particularly those containing hydrophilic moieties other than the charge groups. In such cases, water uptake may no longer correlate with charge density, and the simultaneous percolation of water and charge networks may not occur. The presence of microstructure within the hydrated film may also play a significant role and warrants further investigation.

In addition to examining the interaction between water uptake, conductivity, and IEC, this study aims to clarify the influence of temperature on membrane performance. The apparent activation energies extracted from the temperature dependence of conductivity provide insight into the underlying ionic transport mechanism. They support the idea of a qualitative change in structure, namely the formation of water bridges between charges, which leads to a sharp increase in conductivity.~\cite{Zhongyang_Ge_AEM} Comparing the temperature effect in experiments and simulations also allowed us to validate the physical coupling between the layers of the CLM in circumstances where both layers are independently and simultaneously affected, and far from the condition used for the model's parameterization.  

We found that the effect of temperature on ionic conduction is more complex than expected, as it influences both water uptake in the membrane (which strongly affects conductivity) and the intrinsic resistance of the individual ion pathways, which follows an Arrhenius temperature dependence~\cite{kusoglu_weber_review_PEM}. Although the conductivity measured in our experiments consistently followed Arrhenius behavior, changes in water content in the simulations sometimes gave the appearance of an increased activation energy at higher relative humidity. This occurred because even a small increase in water content near the percolation threshold significantly enhances conductivity (see SI Appendix Figs. 11, 12, and 13). In Figure \ref{fig:temperature_impact} (b), we show that all the previous complexity disappeared once we artificially removed the temperature dependence of water absorption in the simulations. The total activation energy was then fully governed by the dry components at a low water content or the hydrated components at high water content, with a narrow transition near the water percolation threshold.

In summary, we found linking changes in apparent activation energy to shifts in the ionic conduction mechanism in the polymer to be challenging due to the sensitivity of ionic conduction to variations in water content, particularly near the percolation threshold. Furthermore, previous studies have shown that temperature can increase or decrease water absorption, depending on the chemistry of the polymer, underscoring the complexity of predicting its effect~\cite{kusoglu_weber_review_PEM,water_isotherms_dekel, li2010temperature_isotherm}.

The CLM showed compelling agreement with the experimental results, predicting trends in conductivity, water absorption, and their interplay under various conditions. Additionally, the ability of the model to preserve some molecular-level details enabled the examination of local effects, such as the preferential absorption sites in regions of higher local concentration of ions. However, our model has significant limitations that must be considered when interpreting the trends and behaviors it predicts. Some of these limitations are obvious and are justified only by the effort to reduce the complexity of the simulations. For instance, the lattice does not model the polymer's structure; the charges' distribution is determined arbitrarily when the circuit graph is constructed. Moreover, the spatial distribution of charges remains unaffected after the incorporation of water. 

A significant weakness of our model is the dimensionality reduction from a three-dimensional (3D) to a two-dimensional (2D) representation of the system. While this simplification significantly reduces computational complexity and enhances interpretability, it inevitably affects the topology of the resulting water and ion networks. This change in dimensionality strongly raises the percolation threshold observed in our simulations compared to the threshold expected for a four-fold coordinated network in 3D. This difference explains the lower conductivity predicted by our model at low water content compared to the experimental measurements, as shown in SI Appendix Figs 10 (d) and (f). 

Another limitation is the fixed volume of the lattice and the number of hydration sites, leading to isotherms that saturate at high RH, a behavior more typical of solids than polymers~\cite{monson2001Lattice_condensation}. Furthermore, we define the resistance between charges as independent of their separation distance, considering nodes to be disconnected beyond a certain threshold. However, recent work suggests that explicitly incorporating the length and height of the hydration path would improve the agreement with the experimental conductivity~\cite{weber_kusoglu_3D_network}. Finally, neither our model nor the experiments account for entrance effects that may occur at the interface between the membrane and the catalyst or electrolyte, depending on the application, even though these effects can significantly influence the electrical behavior of the membrane~\cite{Janus_entrance}.

In the next stage of this project, we plan to extend our combined experimental and modeling approach to investigate how spatial heterogeneity in charge distribution affects membrane conductivity. Previous studies have identified increased heterogeneity as a potential strategy to enhance ionic transport while reducing water uptake, swelling, and mechanical degradation.~\cite{simulations_morphology,Winey_experiments_simulations_IEC_2,Winey_structure_future,Ion_highways_future,microporosity_structure,wang2024selective_micropores}. 

Finally, despite recent advances~\cite{weaker_electros_water_MM,corti2021structure, QENS_water_ions_dynamics,chuting_water,sujanani2024low_hydration}, a deeper molecular understanding of how water facilitates ion conduction at the single-ion level is still needed. Such knowledge would allow the development of a polymer conduction theory derived from first principles rather than relying on empirical fitting of resistances and activation energies. Key open questions include how water reduces energy barriers for individual ion mobility and how electric resistance depends on charge separation at the single-ion path scale. Targeted molecular dynamics simulations could provide valuable insights by isolating individual ion behavior from collective transport effects within the material.

\section{Conclusion}

This study investigated the effect of Ionic Exchange Capacity (IEC) on Anion Exchange Membranes (AEMs) performance. Our experimental data and simulations revealed an apparently universal relationship between water uptake and ionic conductivity. 

By carefully inspecting the water sorption mechanism, we attributed this behavior to the heterogeneous absorption of water beyond the first hydration shell of the polymer's ionic groups. Water determines the effective IEC of the polymer, which is a measure of how many charges effectively contribute to the conductivity. We also found this heterogeneity to be the reason why the hydration number fails to produce the same convergence of the curves in our experiments. 

We discovered that the network formed by the charges in the polymer and the water hydrogen-bond network must percolate simultaneously. Furthermore, we found that the percolation of a highly conductive path of hydrated charges is determined solely by water content. This important result should hold as long as two conditions are met: first, the charge distribution in the polymer lacks long-distance correlations and second, the ionic groups are the only water-binding sites in the polymer.

In addition, we explored the complex effects of temperature on membrane performance, highlighting the intricate interplay between water absorption and conductivity, and the difficulty in drawing mechanistic insight from apparent activation energy measurements. 

We summarize the mechanism of ionic conduction in hydrated polymers as follows. Local water absorption is driven by the local density of charges within the polymer matrix. When this local density is sufficient, water forms bridges between the charges, creating low-resistance pathways. As the water content reaches the percolation threshold dictated by the water hydrogen bond network, these pathways extend throughout the material, greatly enhancing the conductivity of the membrane.

\section{Methods}

\subsection{Characterization Tools}

\subsubsection{In Situ RH Generator-Ellipsometer-QCM Measurement System}
RH generator, ellipsometer, and quartz crystal microbalance (QCM) were interconnected to simultaneously monitor thickness change and water uptake under controlled humidity conditions. The experimental setup is illustrated in the Supplementary Information Appendix Figure 6. The RH95 humidity generator (Linkam Scientific Instruments) produced humidified gas of specific humidity. A spectroscopic ellipsometer (J.A. Woollam alpha-SE), equipped with a liquid cell connected to inlet and outlet gas tubes, allowed humidified gas from RH generator to flow through. The thin film samples were placed on a stage beneath the liquid cell. Ellipsometer measurements were fitted to the Cauchy layer model to extract thin-film thickness and optical properties. The water absorption mass was measured by the QCM (eQCM 10 M, Gamry Instruments) using a 5 MHz AT cut gold-coated quartz crystal, onto which the thin films were deposited. Temperature control was maintained using a water circulator. The resonant frequency decreased with water absorption as RH increased, which was accurately measured and correlated with the increase in mass using the Sauerbrey equation with a calibration constant of 56.6 Hz $cm^2$ $/mug^{-1}$. A humidity sensor located at the end of the system monitored the humidity level in real time. Data acquisition occurred when the equilibrium between the water chemical potential in the thin films and the water vapor chemical potential in the environment was achieved, as evidenced by the plateau in the QCM frequency signal and stable thickness measurements.

\subsubsection{Fourier Transform Infrared (FTIR) Spectroscopy}
Thin film samples were spin-coated on top of Au-coated Si substrates and subsequently functionalized using a TMA vapor infiltration reaction. Measurements were performed using a Shimadzu IRTracer-100 spectrometer equipped with a diamond prism for attenuated total reflection (ATR). The samples were placed directly on an ATR crystal. Data were collected at room temperature in the range of 400-4000 $cm^{-1}$ with a resolution of 4 $cm^{-1}$.

\subsubsection{Electrochemical Impedance Spectroscopy (EIS)}
Electrochemical impedance spectroscopy (EIS) was performed on top of interdigitated electrodes (IDE) within a humidity chamber (ESPEC SH-242), using a Gamry 600+ potentiostat. Fabrication procedures are described in the Supplementary Information Appendix, Figure 4. The EIS measurements were scanned from 1 MHz to 0.1 Hz under different temperatures and RHs. The ionic resistance data were extracted from the impedance spectrum by fitting it to an equivalent circuit model.

\subsubsection{Local Charge Density Estimation}

The local charge density of a static 2D particle distribution is estimated using Gaussian kernel density estimation (KDE). Given particle positions \(\{\mathbf{x}_i\}_{i=1}^{N}\), KDE approximates the density by centering Gaussian functions at each particle location and summing their contributions:

\begin{equation}
    \hat{\rho}(\mathbf{x}) = \frac{1}{N h^2} \sum_{i=1}^{N} K\left(\frac{\mathbf{x} - \mathbf{x}_i}{h} \right),
\end{equation}

where \(K(\mathbf{u})\) is a two-dimensional Gaussian kernel, and \(h\) is the bandwidth, which controls the spatial resolution. The bandwidth determines how "local" the density estimate is, with larger \(h\) leads to smoother, less localized estimates, while smaller \(h\) provides finer spatial resolution. N is the total number of charges in the network.

The implementation employs \texttt{gaussian\_kde} from \texttt{scipy.stats}, with \(h\) selected based on the correlation length observed in the water sorption probability. The KDE output is scaled such that its integral over the system area equals \(N\), ensuring consistency with the global density \(N/V\), where \(V\) is the total area. The resulting density field is visualized using a color-mapped representation to highlight local charge variations in the lattice.

\subsubsection{Percolation Probability}

Percolation probability is evaluated by analyzing the connectivity of the networks of the CLM. To this end, we employ the Python library \texttt{NetworkX}, using the  \texttt{nx.has\_path} function to determine whether a continuous path exists between two predefined sets of nodes.

For each simulation configuration, we apply this algorithm to the graph constructed from either the charges in the circuit layer or the water sites in the lattice. Percolation is evaluated by verifying whether a spanning path connects opposite edges of the system in both the x and y directions. The percolation probability is then computed as the fraction of configurations in which at least one such path exists, averaged over the entire ensemble of simulated configurations. The percolation threshold corresponds to the water content at which the probability of a spanning path first reaches 50\%.

\section{Supporting Information}
The Supporting Information is available free of charge online.

\section{Notes}\label{sec:Notes}
The authors declare that they have no competing financial interests.

\section{Acknowledgments}\label{sec:Acknowledgments}
This work was supported by the Department of Energy, Office of Basic Energy Sciences, Division of Materials Science and Engineering.
This work used the shared facilities of the Materials Research and Engineering Center at the University of Chicago, supported by the National Science Foundation under award number DMR-2011854.

\end{document}


\maketitle

\clearpage 

\section{Experimental Methods -extended-}

\subsection{Synthesis of Polynorbornene-Based Anion Exchange Films}

Poly(bromo butyl norbornene) (PBBNB) was synthesized from bromo butyl norbornene (BBNB) following a procedure reported previously~\cite{Zhongyang_Ge_AEM}, and is summarized here for completeness:

Poly(bromo butyl norbornene) (PBBNB) was synthesized from bromo butyl norbornene (BBNB) via vinylic addition polymerization in a nitrogen-filled glovebox. A catalyst solution was prepared by mixing (\(\eta^3\)-allyl)Pd(iPr\textsubscript{3}P)Cl (12 mg, 0.034 mmol, Sigma-Aldrich) and lithium tetrakis(pentafluorophenyl)borate$\cdot$(2.5Et\textsubscript{2}O) (Li[FABA], 28 mg, 0.031 mmol, Boulder Scientific Co.) in a 1:1 molar ratio in a solvent mixture of toluene (0.5 g, Sigma-Aldrich, ACS reagent, \(\geq\)99.5\%) and trifluorotoluene (0.5 g, Sigma-Aldrich, 99\%). This mixture was stirred for 20 minutes to generate the cationic palladium catalyst. The BBNB monomer (0.45 g) was purified using three freeze-pump-thaw cycles and then dissolved in toluene (10 mL) to prepare a 5-weight percent solution. The catalyst solution was injected into the monomer solution under vigorous stirring. After polymerization, the product was precipitated three times in methanol (Sigma-Aldrich, ACS reagent, \(\geq\)99.8\%), collected, and dried under vacuum at 60\(^\circ\)C. The resulting PBBNB had a number-average molecular weight (\(M_\text{n}\)) of 68 kg mol\(^{-1}\) and a dispersity of 1.20, as determined by size exclusion chromatography using polystyrene standards.

\subsection{Thin Film Spin-Coating}
For film preparation, PBBNB was dissolved in chlorobenzene at a concentration of 24 mg mL\(^{-1}\) and spin-coated onto various substrates, including interdigitated electrode arrays (IDEs) for impedance spectroscopy, Au-coated Si wafers for FTIR, and Si substrates with 1.5 nm SiO\(_2\) for ellipsometry. Prior to coating, all substrates were cleaned by 2 cycles of ultrasonication in acetone and 2-propanol (each for 5 minutes, Sigma-Aldrich, ACS reagent, \(\geq\)99.5\%). Films were annealed at 80\(^\circ\)C to remove residual solvent. The thickness of the neutral PBBNB film was approximately 60 nm, confirmed by ellipsometry.

\subsection{FTIR Peaks and DoF}
\begin{figure}[H]
    \centering
    \includegraphics[height=0.7\textheight]{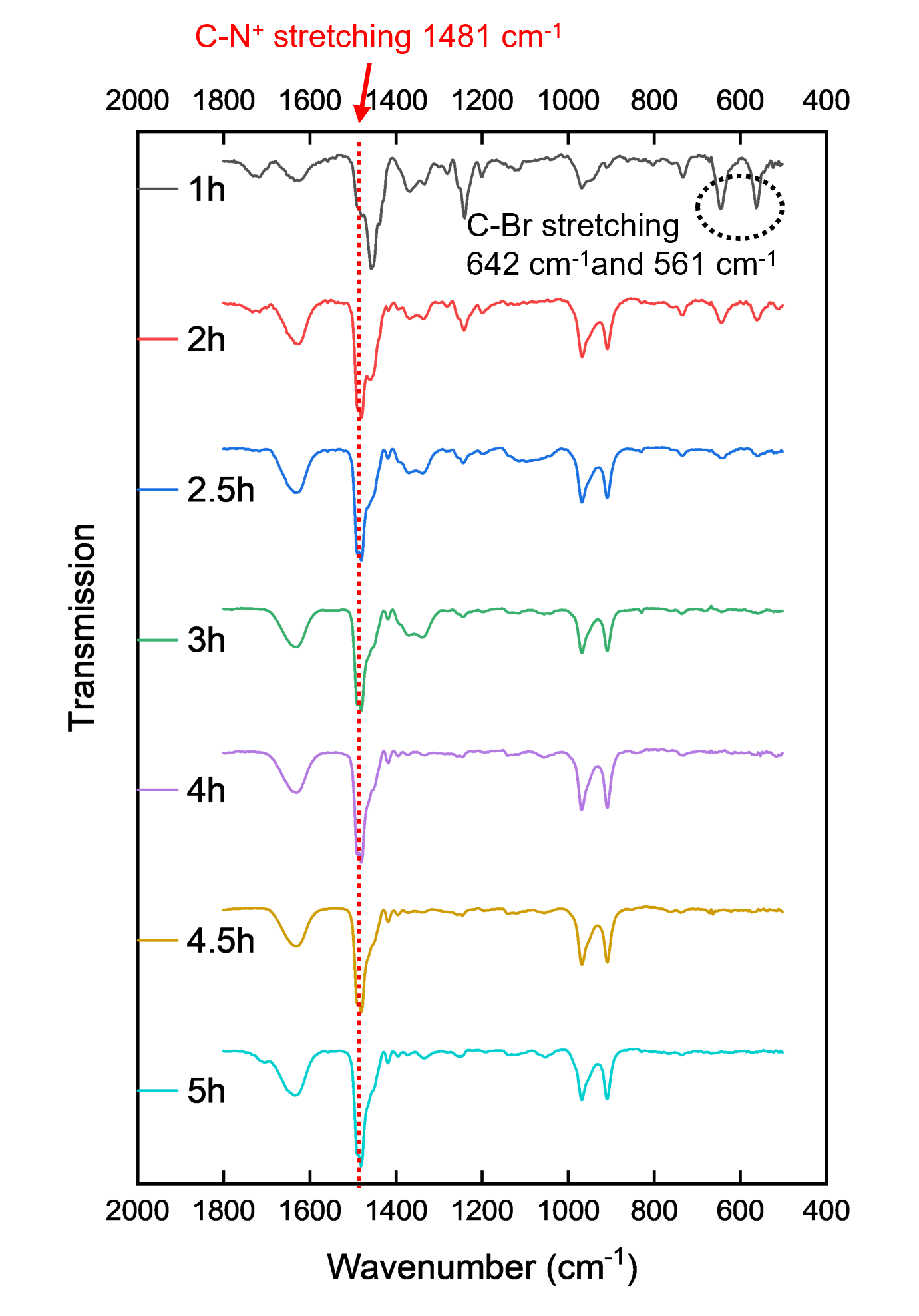}
    \caption{FTIR spectra for different vapor infiltration reaction times in the range of 500 cm\(^{-1}\) to 1800 cm\(^{-1}\). The red dashed line highlights the appearance of the C-N\(^+\) stretching peak at approximately 1481 cm\(^{-1}\). The black dashed circle indicates the disappearance of the C-Br stretching peak at around 642 cm\(^{-1}\) and 561 cm\(^{-1}\).
}

    \label{fig:FTIR peaks and DOF}
\end{figure}

PBBNB films were functionalized by trimethylamine (TMA) to obtain charged PBBNB\(^+\)Br\(^-\) films.  The degree of functionalization, and hence the ion exchange capacity (IEC), was controlled by varying the reaction time.
FTIR was used to measure the degree of functionalization (DoF) in the TMA vapor infiltration reaction as a function of reaction time. The DoF increases as more quaternary ammonium groups replace Br groups, leading to the evolution and disappearance of peaks in the FTIR spectrum. Figure \ref{fig:FTIR peaks and DOF} shows FTIR spectra from 500 cm\(^{-1}\) to 1800 cm\(^{-1}\) for different reaction times. C-Br stretching at the end of the side chain is located at approximately 642 cm\(^{-1}\) and 561 cm\(^{-1}\), while the C-N\(^+\) stretching appears at around 1481 cm\(^{-1}\). As the DoF increases, more C-N\(^+\) bonds start to replace C-Br bonds, resulting in the disappearance of C-Br stretching peaks and the growth of the C-N\(^+\) stretching peak.

\begin{figure}[H]
    \centering
    \includegraphics[width=\textwidth]{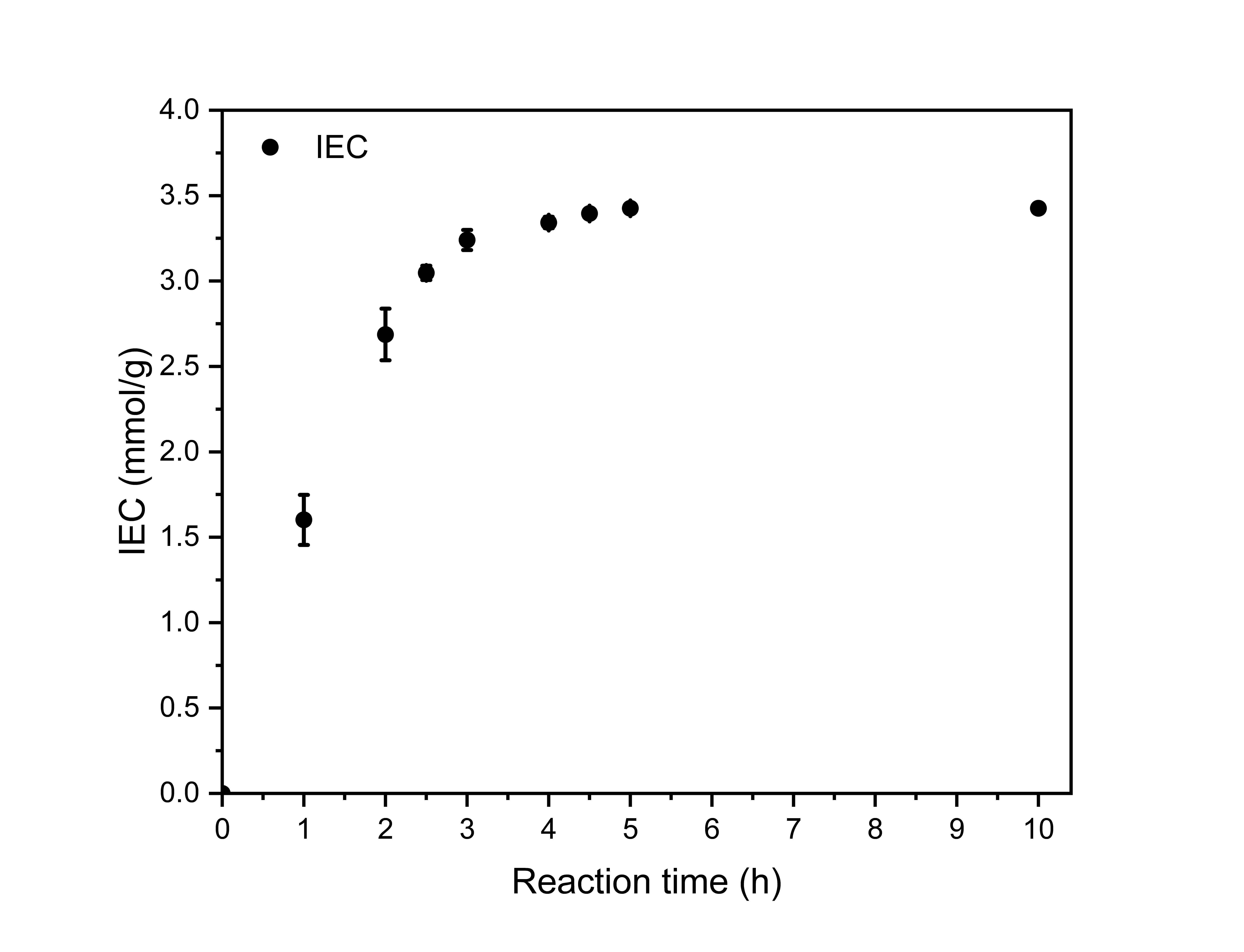}
    \caption{IEC as a function of reaction time. IEC reaches a maximum value of 3.43 mmol/g after 5 hours of reaction. The corresponding DoF is 98
}

    \label{fig:DoF vs time}
\end{figure}

In this study, the degree of functionalization (DoF) is defined as the fraction of C-Br bonds replaced by C-N\(^+\) bonds, calculated using the following equation:
\[
\text{DoF} = \frac{\text{C-N\(^+\) stretching peak area}}{\text{C-N\(^+\) stretching peak area + C-Br stretching peak area}} \times 100\% \quad (1)
\]
The relationship between IEC and DoF can be expressed by the following equation:
\[
\text{IEC} = \frac{\text{DoF} \times 1000}{\text{Average molar mass of repeating unit}} \ \text{mmol/g} \quad (2)
\]
The IEC increases with reaction time, as shown in Figure \ref{fig:DoF vs time}. After 5 hours of reaction time, IEC reaches 3.43 mmol/g and remains unchanged with further extension of the reaction time. Three groups of PBBNB samples were prepared with reaction times of 2 hours, 2.5 hours, and 5 hours. The resulting IECs were 2.69 mmol/g, 3.05 mmol/g, and 3.43 mmol/g, respectively.
After quaternization, film thickness increased to approximately 98 nm, confirmed by ellipsometry.

\begin{figure}[H]
    \centering
    \includegraphics[width=\textwidth]{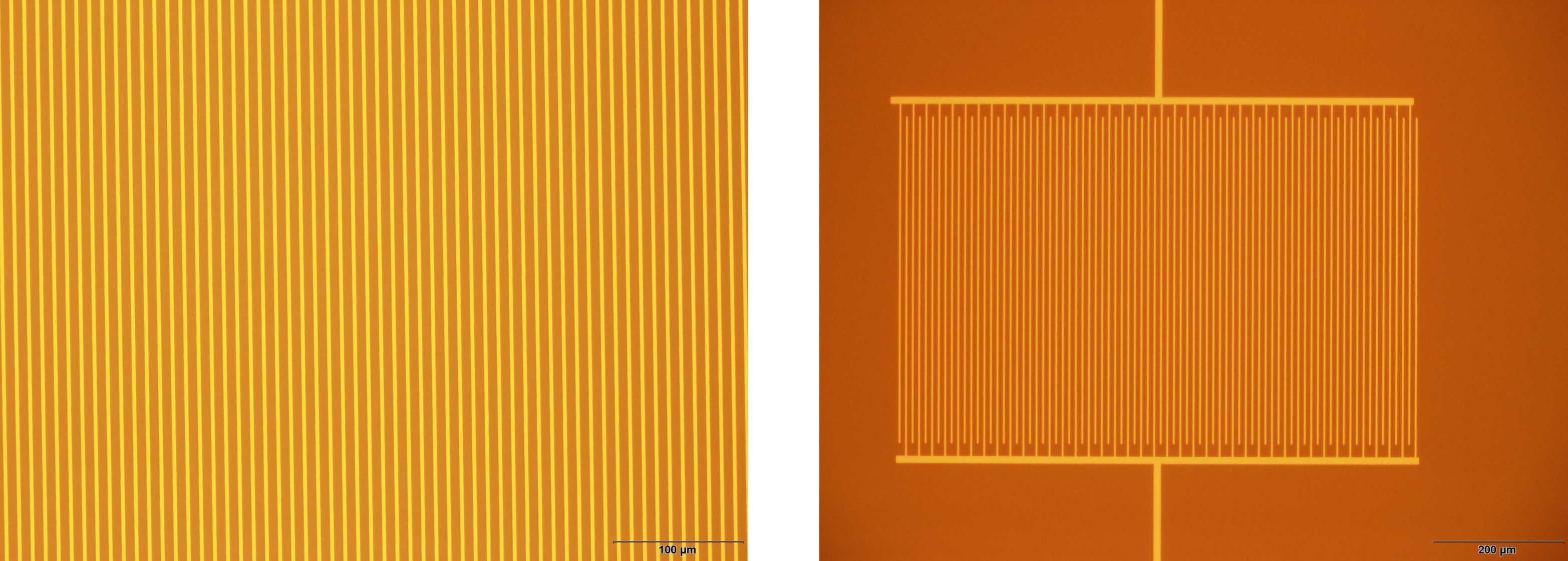}
    \caption{Optical images of interdigitated electrode arrays (IDEs). The bare IDE is shown on the left, and the IDE with the PBBNB\(^+\)Br\(^-\) film on top is shown on the right. The films maintained high quality consistently throughout the characterization process.
}

    \label{fig:microscope}
\end{figure}

\clearpage 

\subsection{Electrochemical Impedance Spectroscopy Measurements}
\begin{figure}[H]
    \centering
    \includegraphics[width=\textwidth]{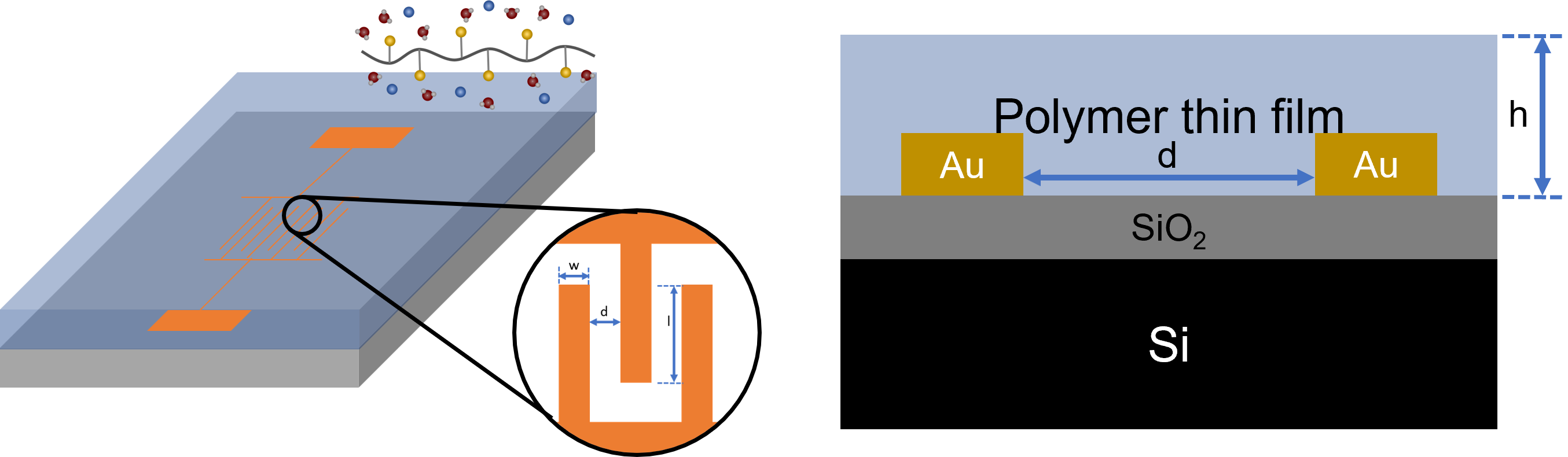}
    \caption{Schematic illustration of the interdigitated electrode (IDE) design and relevant dimensions. 1. Top view 2. Cross-section. The IDE is characterized by four key dimensions: the number of interdigitating electrode teeth (\(N\)), the overlapping length of the electrodes (\(l\)), the width of the electrodes (\(w\)), and the separation distance between electrodes (\(d\)).
}

    \label{fig:IDE}
\end{figure}

Electrochemical impedance spectroscopy (EIS) was performed on top of interdigitated electrodes (IDEs) within a humidity chamber (ESPEC SH-242), utilizing a Gamry 600+ potentiostat. IDEs fabrication details are reported in our previous paper. The top-down and lateral structures are illustrated in Fig \ref{fig:IDE}. 

\begin{figure}[H]
    \centering
    \includegraphics[width=\textwidth]{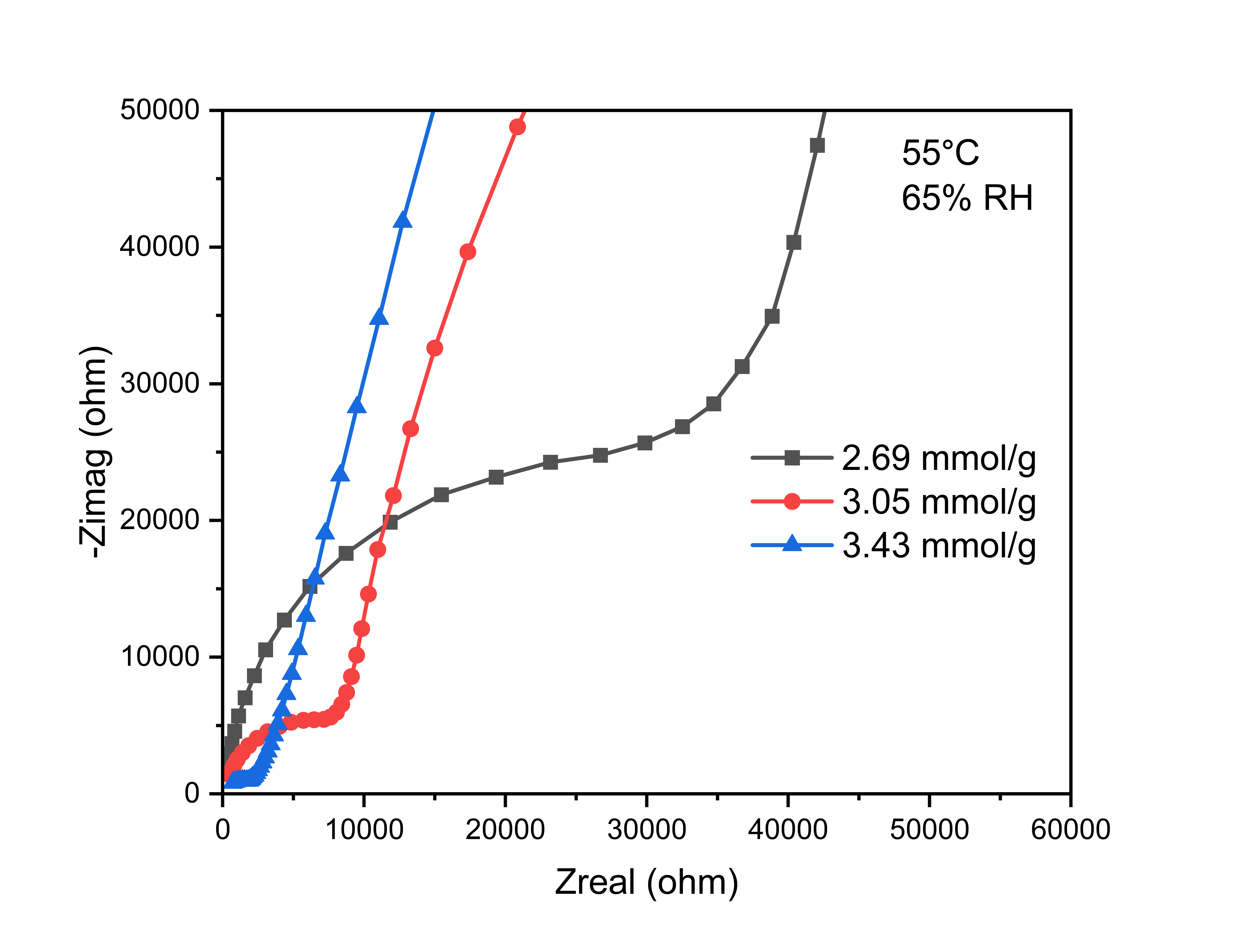}
    \caption{Nyquist plots for three samples at 55°C and 65\% RH with the inserted equivalent circuit model. Black squares represent the sample with an IEC of 2.69 mmol/g, red circles represent the sample with an IEC of 3.05 mmol/g, and blue triangles represent the sample with an IEC of 3.43 mmol/g.
}

    \label{fig:Nyquist}
\end{figure}

First, the open-circuit potential was monitored for fifteen minutes to ensure that water uptake had reached equilibrium under the specified temperature and relative humidity (RH). Subsequently, the complex impedance spectra of the polymer film were measured using a potentiostatic method. During the measurement, a 100 mV AC voltage was applied across the film over a frequency range from 1 Hz to 1 MHz. The resulting impedance spectra were fitted to a physically motivated equivalent circuit model (Figure \ref{fig:Nyquist}) using the Gamry Echem Analyst software with the simplex fitting algorithm.

The ionic resistance data were extracted from the impedance spectrum by fitting it to the equivalent circuit model. The model includes a resistor \(R_f\) and constant phase element \(\text{CPE}_f\) in parallel to describe the thin film response at low and high frequencies, respectively. \(R_f\) refers to the ionic resistance from the film, while \(\text{CPE}_f\) accounts for the non-uniform transport behavior in the high-frequency range. The constant phase element \(\text{CPE}_{\text{int}}\) in series with the film components corresponds to the "imperfect" capacitor-like behavior near the ion-blocking electrodes, including the formation of the electric double layer. The circuit also includes \(R_s\) to account for resistive losses from the experimental setup and a capacitor \(C_{\text{sub}}\) to describe the capacitance of the silicon dioxide substrate. Figure \ref{fig:Nyquist} shows the Nyquist plots for three samples at 55\(^\circ\)C and 65\% RH with the inserted equivalent circuit model. The Nyquist plots show a partial semicircle followed by a diffusion tail, where the diameter for the semicircle corresponds to the ionic resistance \(R_f\) in the film. The sample with high IEC exhibited a smaller diameter, indicating a smaller ionic resistance \(R_f\) under the same conditions.

The film resistance \(R_f\) obtained by this method was then used to determine the ionic conductivity (\(\sigma\)) using equation 3.
\[
\sigma = \frac{1}{R_f} \frac{d}{l(N-1)h} \quad \quad \quad \quad \quad \quad \quad \quad \quad \quad (3)
\]
In the equation, \(d = 8 \ \mu\text{m}\) is the distance between adjacent electrode teeth, \(l = 500 \ \mu\text{m}\) is the effective electrode length, \(N = 80\) is the number of electrode teeth, and \(h\) is the thickness of the film.

The ionic conductivity reported in the main text represents the average of at least three samples, with error bars indicating the standard deviation. 

\clearpage 

\subsection{\textit{in situ} RH Generator-Ellipsometer-QCM Measurement System}
Typically, AEM fuel cell studies focus on micrometer-scale membranes using standard characterization tools and procedures that are not applicable to our nanometer-scale thin films. Consequently, we employ different measurement platforms to gain insights from a smaller perspective. This approach helps bridge the knowledge gap between the behavior of nanoscale ionomers and larger-scale membrane studies, offering a more comprehensive understanding of polyelectrolyte performance in practical fuel cell applications.

\begin{figure}[H]
    \centering
    \includegraphics[width=\textwidth]{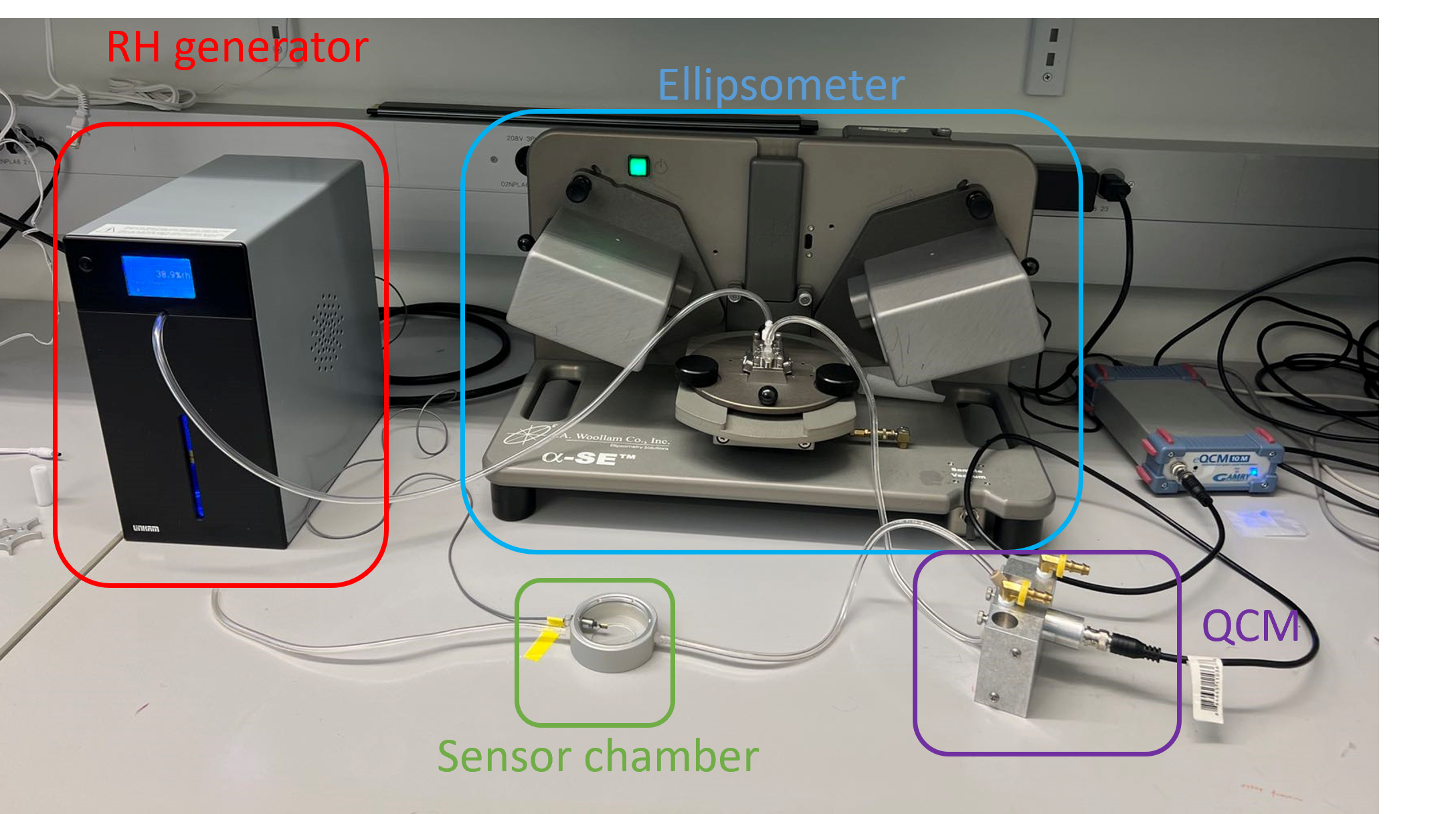}
    \caption{Picture of the \textit{in situ} RH generator-Ellipsometer-QCM measurement system for accessing thin film thickness and water uptake as functions of RH. The setup enables the controlled flow of humidified gas through a sequential pathway: from the RH generator to the ellipsometer, where the sample is covered by a liquid cell, then to the QCM, and finally to the sensor chamber before dispersing into the ambient environment
}

    \label{fig:in situ system}
\end{figure}

RH generator, ellipsometer and quartz crystal microbalance (QCM) were interconnected to monitor the thickness change and water uptake simultaneously (WU) under controlled humidity conditions at room temperature. The \textit{in situ} RH generator-Ellipsometer-QCM measurement system is illustrated in Figure \ref{fig:in situ system}. The system includes several key components arranged in sequence to ensure precise control and measurement. The setup is as follows:

1.	RH Generator: The RH95 humidity generator (Linkam Scientific Instruments) is used to produce a gas stream with controlled humidity levels. This gas stream is directed through the system to simulate different environmental conditions.

2.	Ellipsometer: The humidified gas first flows through the ellipsometer (J.A. Woollam alpha-SE), where the sample is positioned under a liquid cell. The liquid cell has inlet and outlet gas tubes, allowing the humidified gas to flow over the sample. The ellipsometer measures the thickness and optical properties of the thin film in real time as the humidity changes. Ellipsometer measurements were fitted to the Cauchy layer model to extract thin film thickness and optical properties.

3.	QCM (Quartz Crystal Microbalance): After passing through the ellipsometer, the gas continues to the QCM (eQCM 10M, Gamry Instruments). The QCM is equipped with a 5 MHz AT-cut gold-coated quartz crystal where the thin films are deposited. This component measures the mass of water absorbed by the film based on the frequency change of the quartz crystal. Stable room temperature was maintained using a water circulator.  

4.	Sensor Chamber: Following the QCM, the gas enters the sensor chamber, where a humidity sensor monitors the real-time RH levels to ensure accuracy and consistency throughout the measurement process.

5.	Exhaust to Ambient: Finally, the gas exits the system and is released to the ambient environment.

This setup allows for simultaneous monitoring of thin film thickness and water uptake under controlled humidity conditions, providing comprehensive data on the film's response to varying RH levels. Data acquisition occurred when equilibrium between water chemical potential in the thin films and water vapor chemical potential in the environment was achieved, as evidenced by the plateau in the QCM frequency signal and stable thickness measurements.

In the water uptake (WU) calculation Equation (5), the resonant frequency decreases with water absorption as the relative humidity (RH) increases. This decrease is accurately measured and correlated to mass increase using the Sauerbrey equation, with a calibration constant of \( 56.6 \, \text{Hz} \, \text{cm}^2 \, \mu\text{g}^{-1} \). Here, \( f_0 \) refers to the frequency in the dry state, and \( f_n \) refers to the frequency at a specific RH. The secondary constant of \( 4.24 \, \text{ng} \, \text{Hz}^{-1} \) is derived by canceling out the effective area \( (0.24 \, \text{cm}^2) \) in the calibration constant. In the denominator of Equation (5), the dry mass of the thin film is calculated using the density and volume. The effective area, \( S \), is \( 0.24 \, \text{cm}^2 \) in the experiment. \( t_0 \) is the thickness in the dry state, which can be obtained from the thickness measurements. The density of the dry polymer, \( \rho_{\text{dry}} \), is estimated based on the density of PBBNB and the thickness expansion after the functionalization process. The density of PBBNB was measured to be \( 1.291 \, \text{g/cm}^3 \) using a density determination kit with Excellence XP/XS analytical balances (Mettler Toledo\texttrademark). The densities of dry PBBNB\(^+\)Br\(^-\) were \( 1.161 \, \text{g/cm}^3 \) (for IEC = 2.69 mmol/g), \( 1.148 \, \text{g/cm}^3 \) (for IEC = 3.05 mmol/g), and \( 1.126 \, \text{g/cm}^3 \) (for IEC = 3.43 mmol/g).

\[
WU = \frac{M_w - M_d}{M_d} = \frac{(f_0 - f_n) \cdot 4.24}{\rho_{\text{dry}} \cdot S \cdot t_0} \quad \quad \quad \quad \quad \quad \quad \quad \quad \quad \quad \quad (5)
\]

The hydration number (\(\lambda\)), representing the number of water molecules per ionic group, can be calculated from WU and IEC using equation (6). 
\[
\lambda = \frac{1000}{18.02} \frac{WU}{IEC} \quad \quad \quad \quad \quad \quad \quad \quad \quad \quad \quad \quad (6)
\]

\clearpage 

\section{Polymer Swelling and Water Uptake}

\begin{figure}[H]
    \centering
    \includegraphics[width=\textwidth]{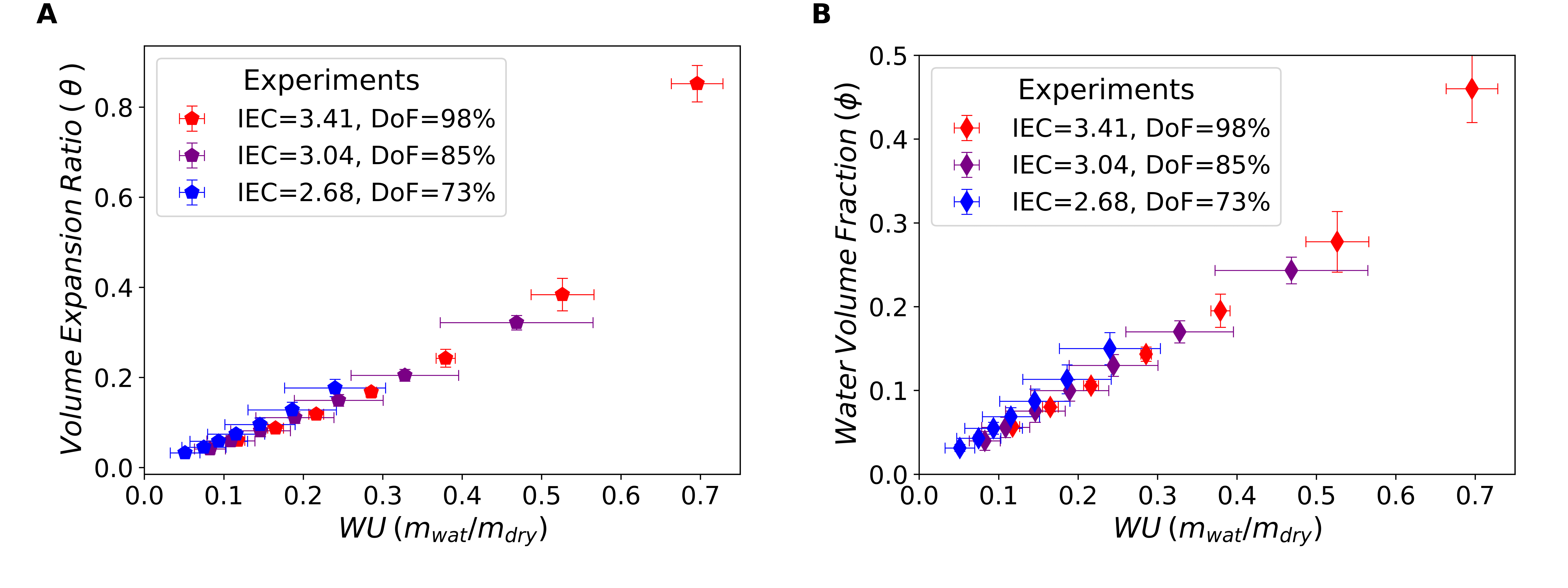}
    \caption{Polymer thickness expansion and water volume fraction. (A) Film thickness expansion as a function of water content. The expansion ratio $\theta$ is defined as the ratio between the absorbed water volume and the dry polymer volume, i.e., $\theta = V_{\mathrm{water}} / V_{\mathrm{dry\,polymer}}$. (B) Water volume fraction $\phi$ calculated from (A) using the relation $\phi = \theta / (1 + \theta)$, which assumes additive volumes and negligible excess volume of mixing.}

    \label{fig:volume}
\end{figure}

Figure~\ref{fig:volume} shows the polymer thickness expansion as a function of water uptake. As the film absorbs more water at higher relative humidity (RH), its thickness increases, with more pronounced expansion observed at higher ion-exchange capacities (IEC). Figures~\ref{fig:volume}(a) and (b) show that both the expansion ratio and the corresponding water volume fraction collapse onto a single curve. At high RH and IEC, surface roughness on the gas-facing side of the film affects thickness detection by diffuse reflection, likely explaining the deviation of the final point in Figure~\ref{fig:volume}(a).

\clearpage 

\section{Effect of Chosen Substrate and Initial Film Thickness on Water Uptake}

\begin{figure}[H]
    \centering
    \includegraphics[width=\textwidth]{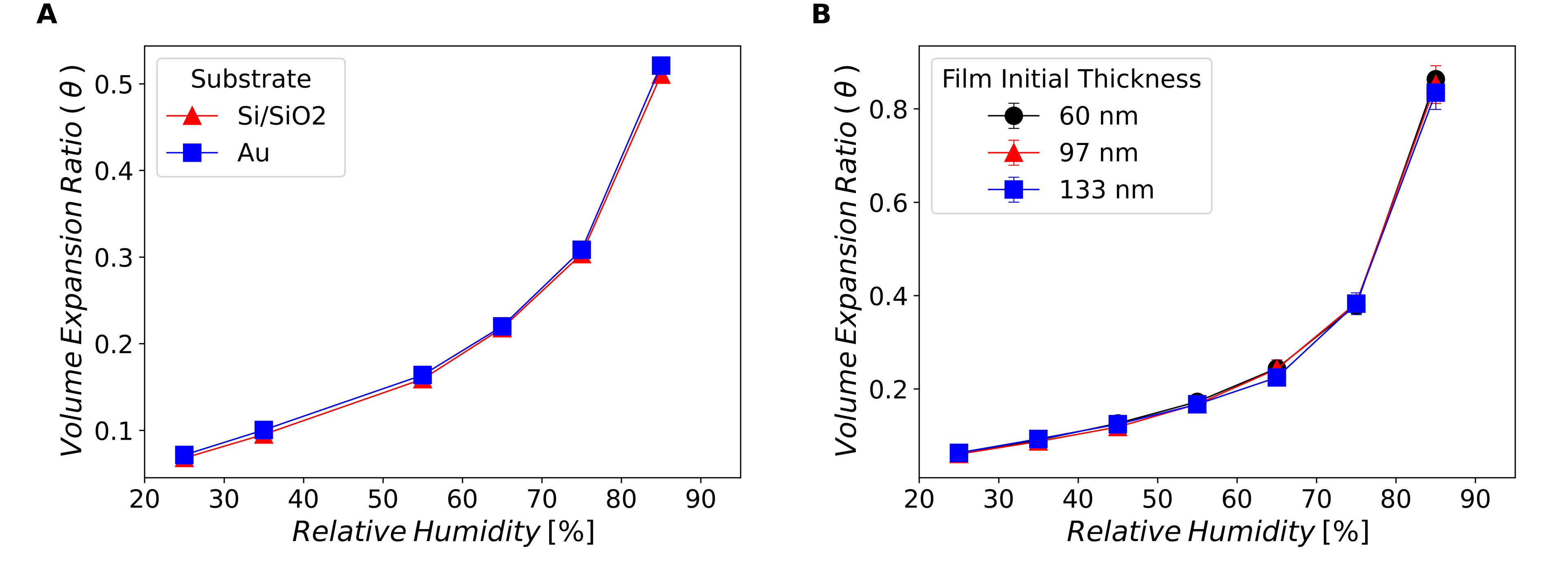}
    \caption{Swelling behavior of polymer films prepared under different conditions. (A) Volume expansion ratio as a function of relative humidity measured by ellipsometry for films deposited on Si/\( \mathrm{SiO}_2 \) and gold substrates. While the IEC of these samples was not independently confirmed by FTIR, the vapor infiltration time (4~h) was controlled to ensure equal ion exchange capacity (approximately 3.2~mmol/g). (B) Volume expansion ratio as a function of relative humidity for polymers with different initial thicknesses after functionalization. The 97~nm film corresponds to the sample used throughout the main text. These samples were fully functionalized after 5~h of vapor infiltration, achieving an IEC of 3.4~mmol/g. All measurements were performed at approximately 25~\textdegree C.}

    \label{fig:substrates}
\end{figure}

Figure~\ref{fig:substrates} shows the volume expansion ratio of polymer films under different preparation conditions. To accommodate various measurement techniques, the polymer was deposited on different substrates: interdigitated electrodes for conductivity, Au-coated Si wafers for FTIR, and Si/\( \mathrm{SiO}_2 \) substrates for ellipsometry. Despite these variations, the data in Figure~\ref{fig:substrates}(a) show that the swelling behavior is not significantly affected by the substrate. In addition, Figure~\ref{fig:substrates} (b) confirms that films with initial dry thicknesses of approximately 60, 97, and 133~nm exhibit similar swelling responses. Recalling the direct proportionality between volume expansion and water uptake established in Figure~\ref{fig:volume}, these results demonstrate the trends presented in the main text are robust with respect to both substrate and film thickness.

\clearpage 

\section{Comparison of Different Water Content Descriptors for Normalizing Ionic Conductivity}

\captionsetup{format=plain, belowskip=0pt, aboveskip=10pt}

\begin{center}
  \includegraphics[height=0.8\textheight]{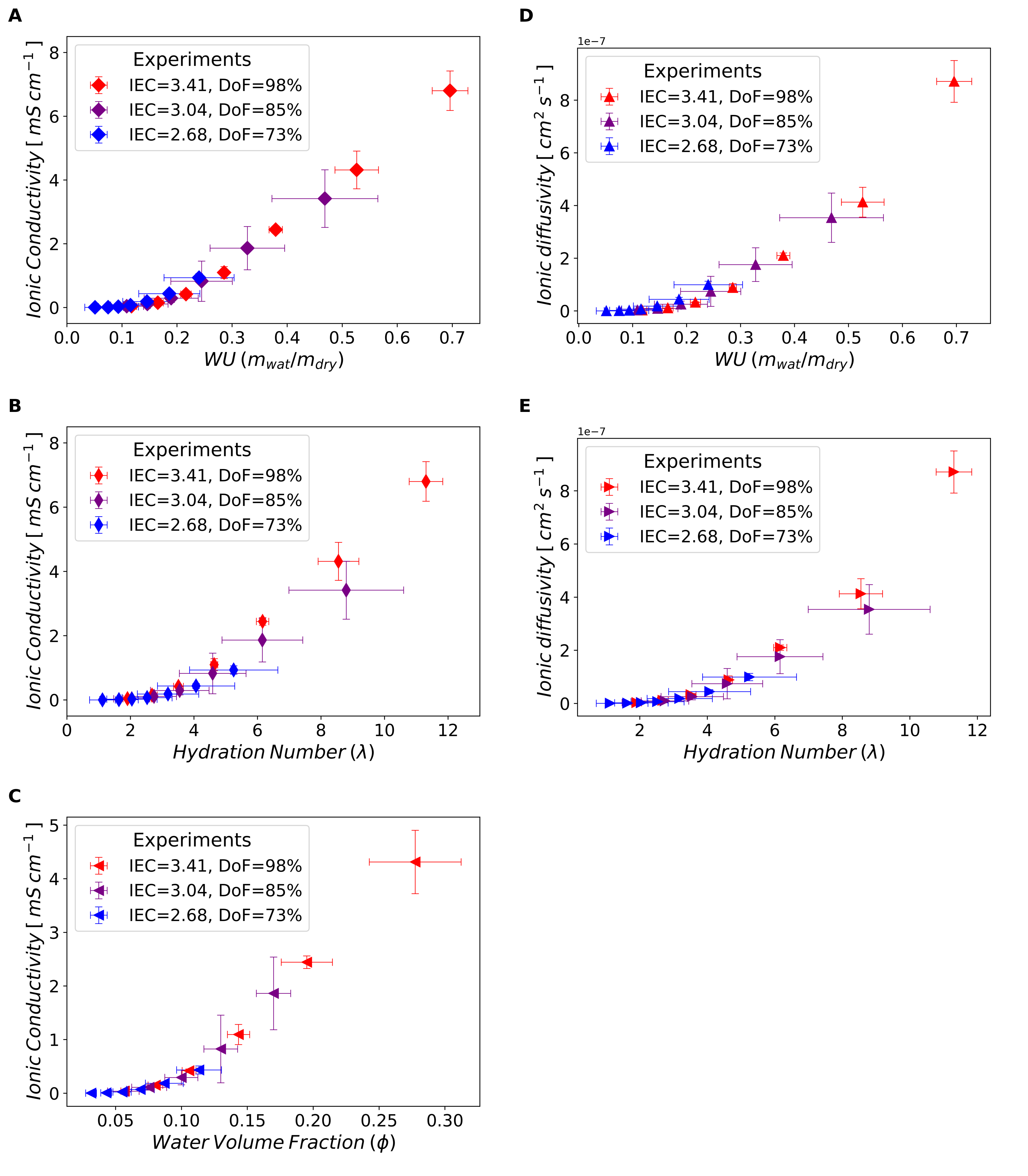}
\end{center}

\captionof{figure}{
Comparison of different water content descriptors and their ability to unify the behavior of samples with different ion-exchange capacities (IECs). (A) and (D) show the ionic conductivity and diffusivity as a function of water content, the primary control parameter used in the main text. (B) and (E) plot the same transport properties against the hydration number (number of water molecules per charged group); in this case, the curves from different IECs do not align as well as before. (C) shows that using the water volume fraction as a descriptor improves the alignment of conductivity data across IECs. The point corresponding to the highest water content was omitted from panel (C) due to anomalous behavior (see SI Appendix, Figure~\ref{fig:volume}). Diffusivity in (D) and (E) was calculated from the measured ionic conductivity using the Nernst–Einstein relation: $D = \frac{\sigma k_B T}{n z^2 e^2}$, where $\sigma$ is the ionic conductivity, $n$ is the number density of mobile ions, $z$ their valence, $e$ the elementary charge, $k_B$ the Boltzmann constant, and $T$ the temperature.
}
\label{fig:collapses}

\section{Percolation in Two and Three Dimensions and Water Content}

\begin{center}
  \includegraphics[height=0.8\textheight]{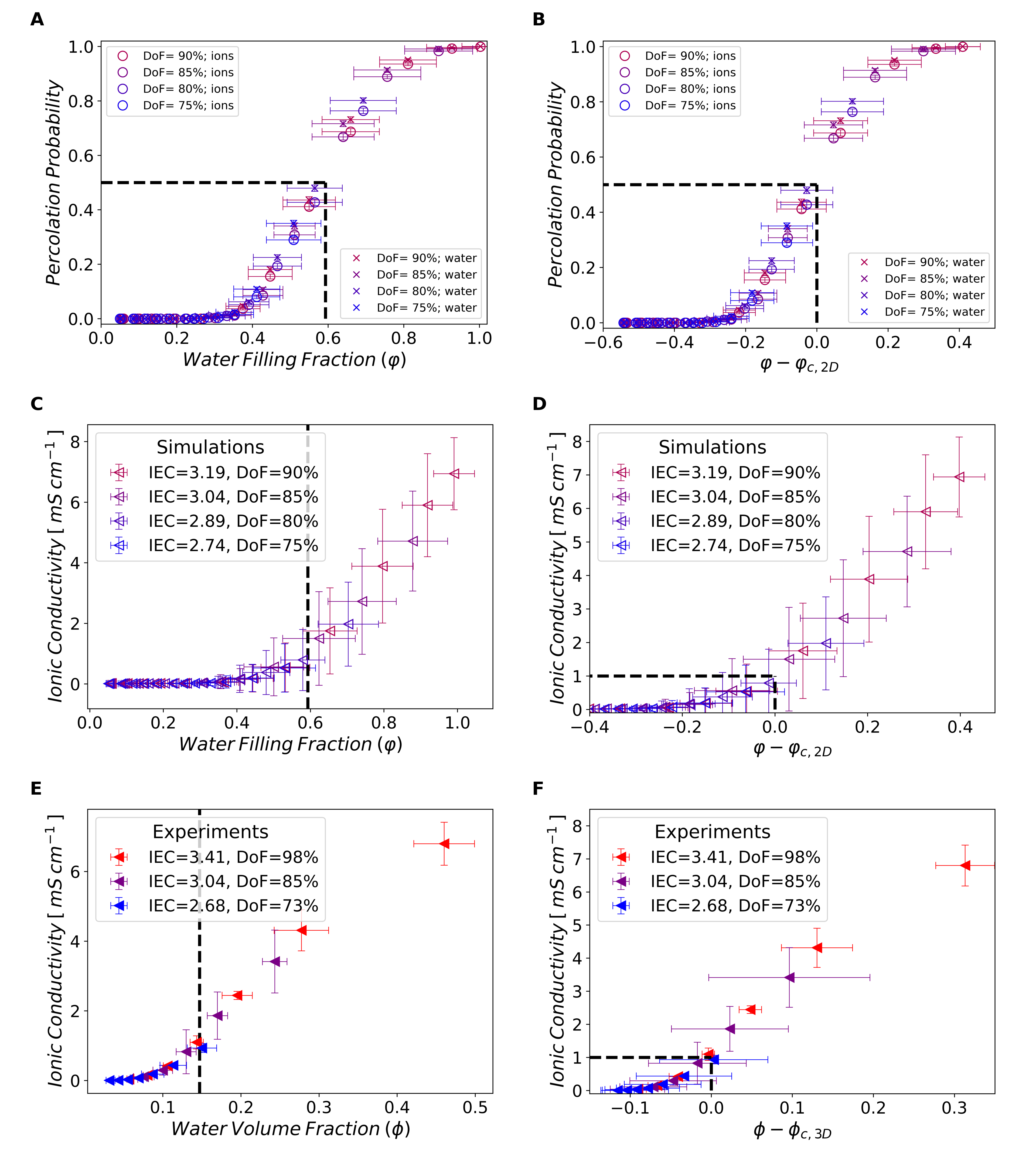} 
\end{center}

\captionsetup{format=plain, belowskip=0pt, aboveskip=10pt}

\captionof{figure}{%
Percolation threshold in 2D and 3D systems: comparison between simulations and experiments.
(A) Percolation probability as a function of water filling fraction in 2D lattice simulations. The filling fraction refers to the fraction of occupied lattice sites (i.e., the site occupancy probability), which is the standard metric for defining the percolation threshold in 2D lattice models. Note that in this case, the filling fraction does not correspond to the water volume fraction in the polymer. The observed percolation threshold agrees with the expected value for a 2D square lattice (dashed line, $P_c = 0.593$).\cite{mertens2022percolation_treshold_square}
(B) Percolation probability scaled by the expected threshold, clearly illustrating that the probability of percolation is nearly exactly 0.5 at the theoretical percolation threshold.
(C) Simulated ionic conductivity as a function of filling fraction, illustrating the onset of fast ion transport near the percolation threshold.
(D) Filling fraction scaled by the theoretical percolation threshold in 2D. Dashed lines indicate the conductivity at the threshold.
(E) Experimental ionic conductivity as a function of water volume fraction in hydrated polymer films. Here, the water volume fraction corresponds directly to the filling fraction in the material. The dashed line indicates the expected percolation threshold for a fully connected hydrogen-bond network in a tetrahedral lattice ($P_c = 0.147$).\cite{frisch1961percolation_treshold_ice}
(F) Ionic conductivity as a function of scaled filling fraction; dashed lines mark the conductivity at the percolation threshold. The comparison between panels (D) and (F) shows that the discrepancy in conductivity at low water content between experiments and simulations (CLM) disappears when accounting for the dimensionality-dependent shift in percolation probability. Overall, these results indicate that in the CLM, conductivity is governed by the percolation of the water network and is independent of the polymer matrix. In the real polymer, the sharp increase in conductivity near the expected 3D percolation threshold suggests that water percolation might also be the dominant mechanism governing charge transport.
}

\label{fig:percolation}

\clearpage 

\section{Temperature Effect on Conductivity}

\subsection{Experiments}

\begin{figure}[H]
    \centering
    \includegraphics[width=\textwidth]{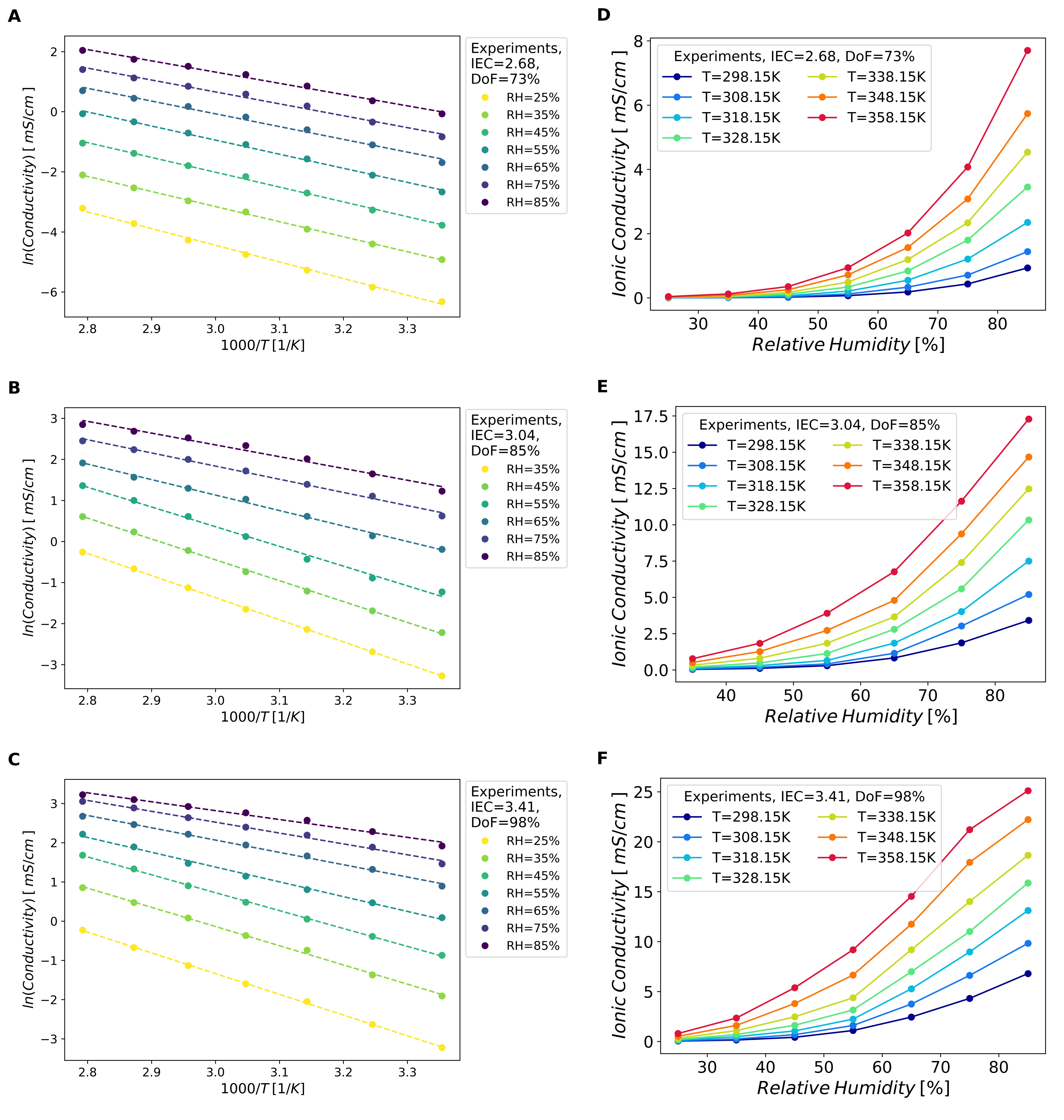}
    \caption{Temperature effect on conductivity: Experiments. (A), (B) and (C) show Arrhenius-like plots for the polymer of 2.68,3.04, and 3.41 mmol/g IEC, respectively. (D), (E) and (F) display curves of conductivity vs RH at all the temperatures used to construct the Arrhenius plots on the left column.    }

    \label{fig:Temp_conductivity_exp}
\end{figure}

\clearpage 

\subsection{Simulations}

\begin{figure}[H]
    \centering
    \includegraphics[height=0.7\textheight]{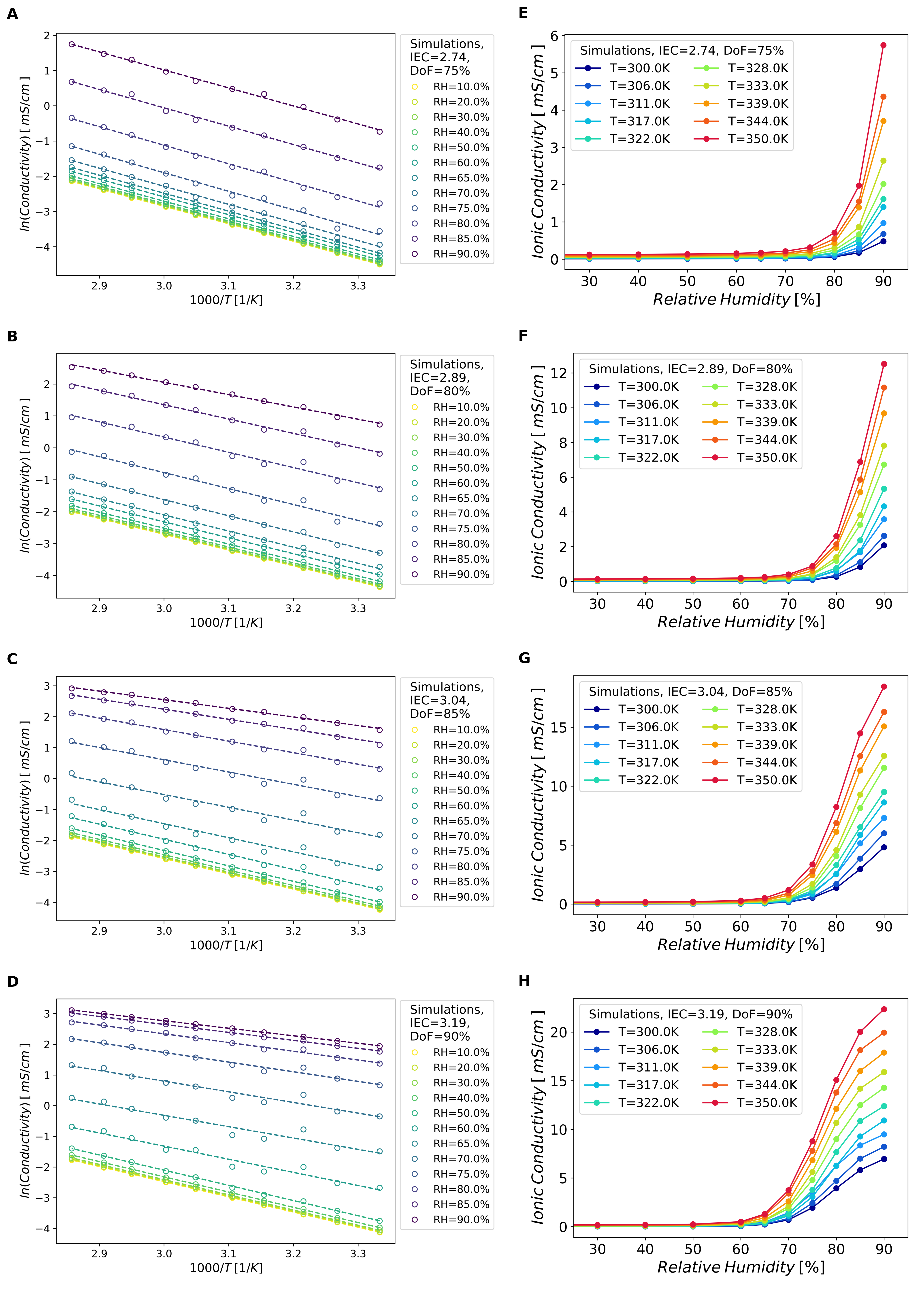}
    \caption{Temperature effect on conductivity: simulations. (A), (B), (C), and (D) show Arrhenius-like plots for the polymer of 2.74, 2.89, 3.04, and 3.19 mmol/g IEC, respectively. (D), (E), (F), and (G) display curves of conductivity vs RH at all the temperatures used to construct the Arrhenius plots on the left column.    }

    \label{fig:Temp_conductivity_simulations}
\end{figure}

\clearpage 

\section{Temperature Effect on Water Sorption}

\begin{figure}[H]
    \centering
    \includegraphics[height=0.7\textheight]{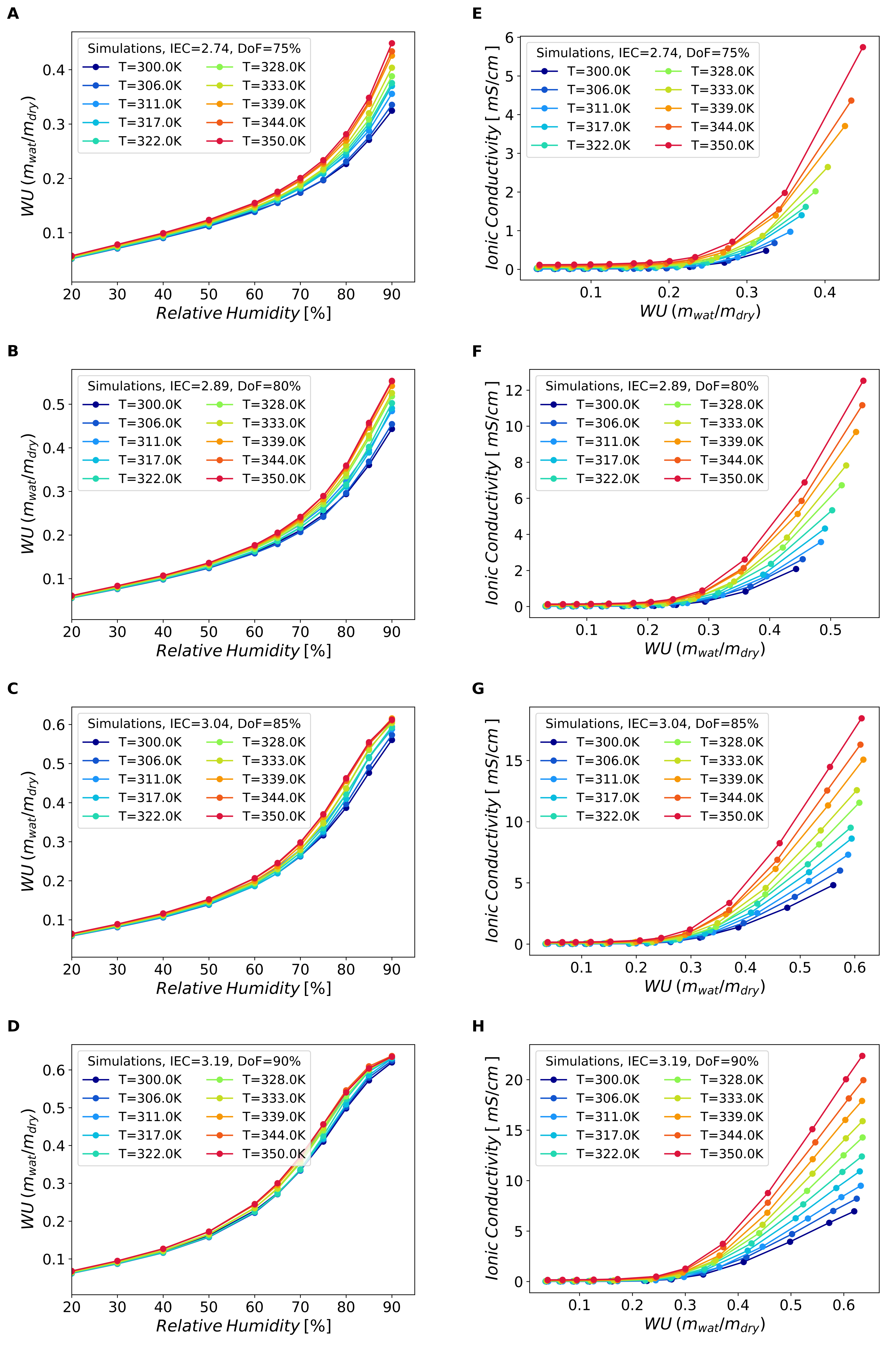}
    \caption{Temperature effect on water adsorption: Simulations. Panels (A), (B), (C), and (D) display the curves of water uptake vs. relative humidity (RH) at all temperatures used for the Arrhenius plots for polymers with IEC values of 2.74, 2.89, 3.04, and 3.19 mmol/g, respectively. In panels (E), (F), (G), and (H), we demonstrate the impact of these differences in adsorption by plotting ionic conductivity against water uptake. It should be noted that at lower IECs, the temperature has a larger-than-normal effect because it pushes the water adsorption across the percolation threshold.   }

    \label{fig:Temp_WU_simulations}
\end{figure}

\clearpage 

\section{Hysteresis and Metastability}
\begin{figure}[H]
    \centering
    \includegraphics[width=0.5\textwidth]{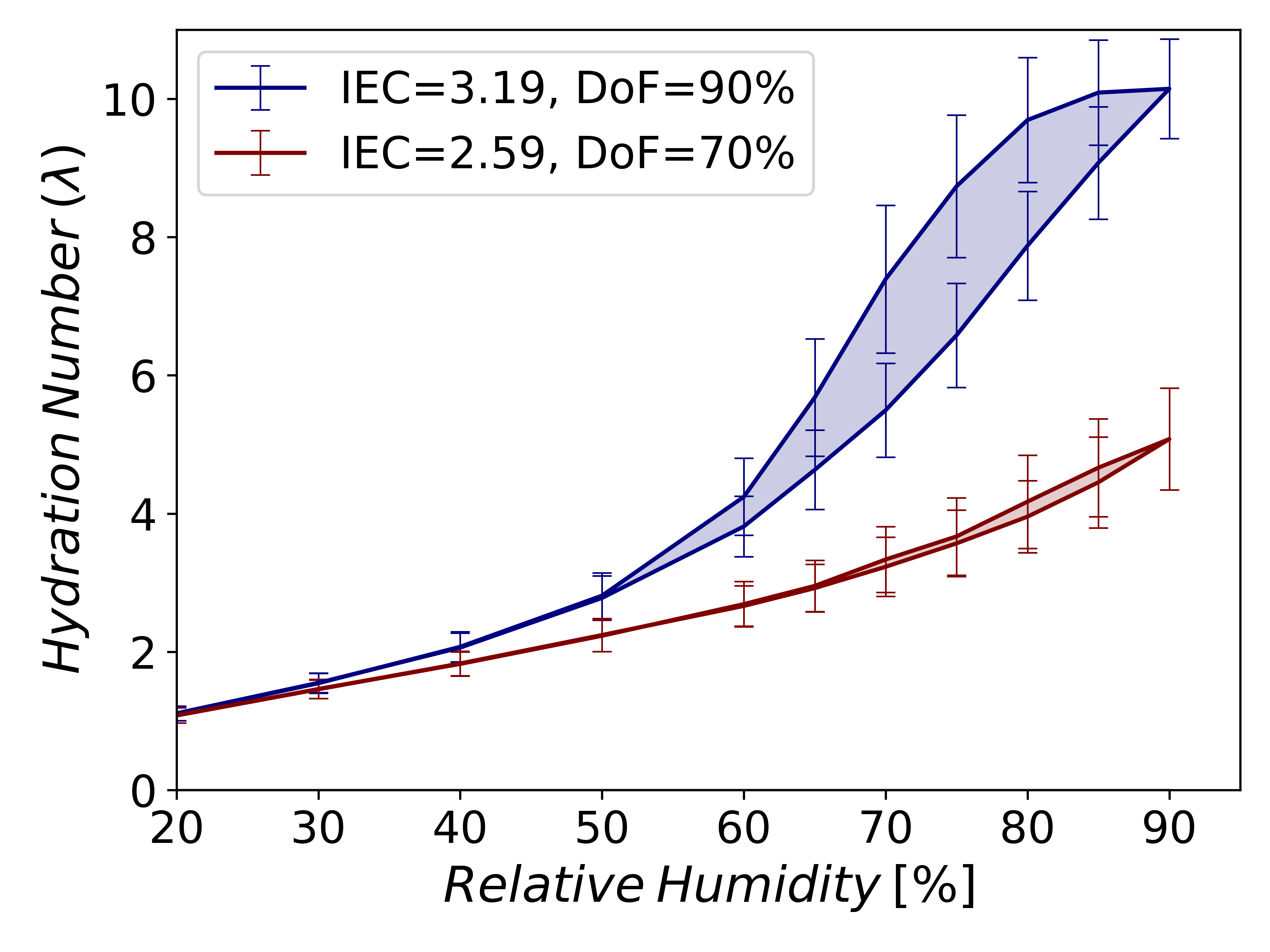}
    \caption{Hysteresis of water adsorption in simulations. The figure shows the adsorption curves followed by desorption using the same number of equilibration steps. It can be seen that the desorption curves retain part of the water content until lower RH. The water in the shaded area is metastable and corresponds to water adsorbed beyond the ion's first hydration shell. In polymers with lower IEC, the hysteresis effect is minimal due to the lack of this type of water. It is important to note that hysteresis occurs because the system does not reach chemical equilibrium within the simulated Monte Carlo steps. Extending the simulation for more steps would eventually cause the two curves to coincide.      }

    \label{fig:hysteresis}
\end{figure}

\clearpage 

\section{Water Model Phase Behavior}

\begin{figure}[H]
    \centering
    \includegraphics[height=0.7\textheight]{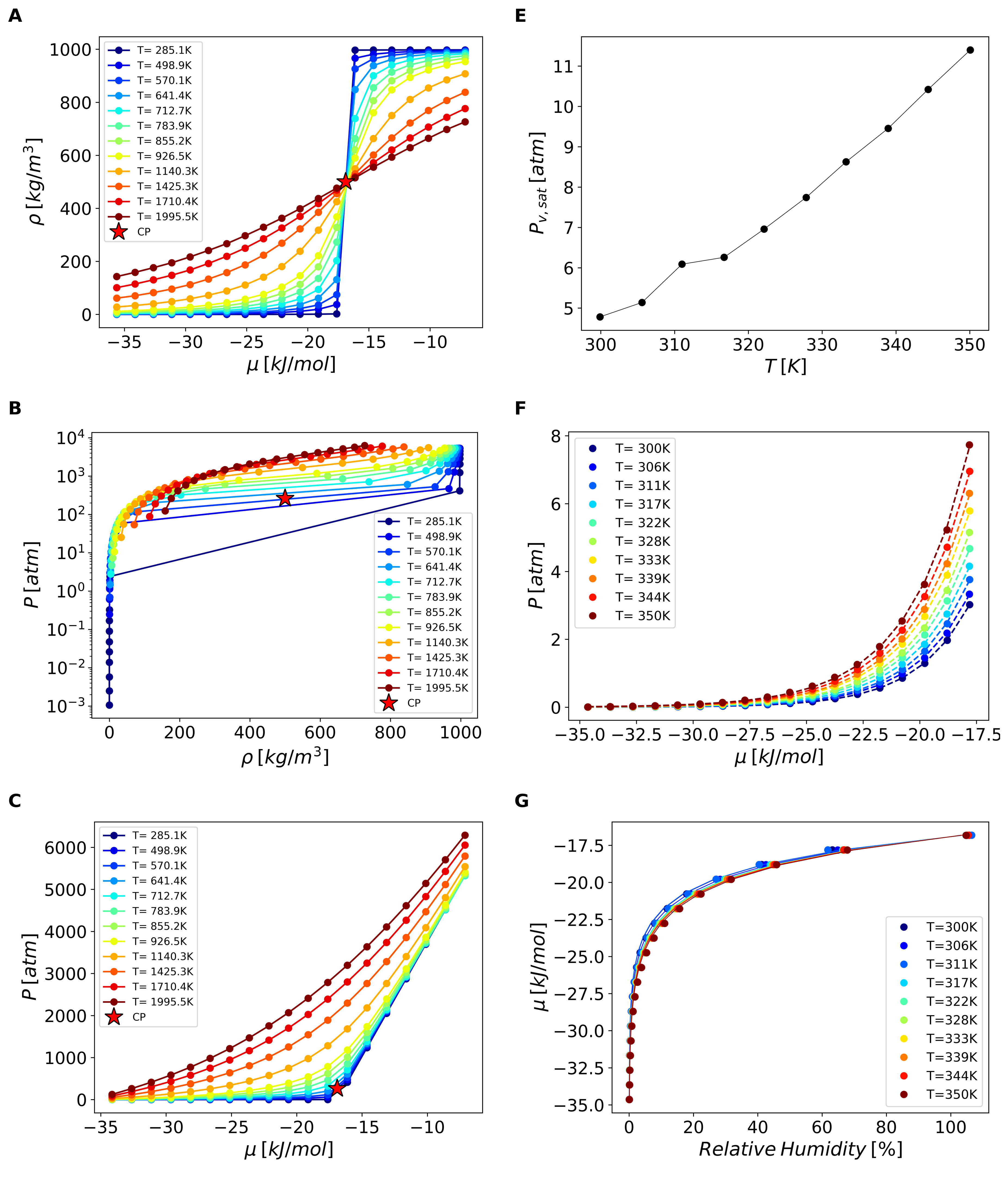}
    \caption{Pure solvent characterization. (A) Density vs chemical potential for various temperatures. (B) Pressure vs density. (C) Pressure vs chemical potential. (D) Saturation pressure for the fully parametrized water model vs temperature. (E) Vapor pressure vs chemical potential. (F) Chemical potential vs relative humidity of the pure solvent. This last plot illustrates the chemical potential that is used in the grand canonical simulations of water adsorption. The red stars indicate the model's critical point.  }

    \label{fig:solvent}
\end{figure}

\clearpage 

\bibliography{IEM_references} 